\documentclass[preprint,showpacs,preprintnumbers,amssymb,superscriptaddress,aps,prd,nofootinbib,11pt]{revtex4-1}

\selectfont

\usepackage{graphicx}       
\usepackage{dcolumn}        
\usepackage{bm}             
\usepackage{psfrag}
\usepackage{tensor}
\usepackage[usenames]{color}
\definecolor{navyblue}{rgb}{0.0, 0.0, 0.5}
\usepackage[linktocpage,colorlinks=true,allcolors=navyblue]{hyperref}
\usepackage{amsmath}
\usepackage{subfigure}

\newcommand{\be}{\begin{equation}}
\newcommand{\ee}{\end{equation}}

\newcommand{\beq}{\begin{equation}}
\newcommand{\eeq}{\end{equation}}

\begin{document}

\preprint{}

\title{Wave focusing by submerged islands and gravitational analogues}

\author{Theo Torres}
 \email{theo.torres_vicente@kcl.ac.uk}
\affiliation{Department of Physics, King's College London, The Strand, London WC2R 2LS, UK}
\author{Max Lloyd}
\affiliation{Department of Applied Mathematics and Theoretical Physics, University of Cambridge, CB3 0WA, UK}
\author{Sam R. Dolan}
\affiliation{Consortium for Fundamental Physics,
School of Mathematics and Statistics,
University of Sheffield, Hicks Building, Hounsfield Road, Sheffield S3 7RH, UK}
\author{Silke Weinfurtner}
\affiliation{School of Mathematical Sciences, University of Nottingham, University Park, Nottingham, NG7 2RD, UK}

\date{\today}

\begin{abstract}
We study water waves propagating over a smooth obstacle in a fluid of varying depth, motivated by the observation that submerged islands in the ocean act as effective lenses that increase the amplitude and destructive power of tsunami waves near focal points. 
We show that islands of substantial height (compared to the water depth) lead to strong focusing in their immediate vicinity, and generate caustics of either cusp or butterfly type. We highlight similarities and differences with focusing of (high-frequency) gravitational waves by a neutron star. 
In the linear regime, the comparison is made precise through an effective-spacetime description of the island-fluid system.
This description is then put to practical use: we identify caustics by solving the Raychaudhuri equation (a transport equation) along rays of the effective metric. 
Next, the island-fluid scattering processes are examined in detail (i.e.~deflection angle, phase shifts, scattering amplitudes) using numerical simulations and analytical techniques, including the eikonal approximation and its generalisation in the form of the Gaussian beam approximation. We show that the techniques capture the key features of the simulations. 
Finally, we extend the eikonal approximation to the dispersive regime, demonstrating that the essential features are robust in dispersive settings. This paves the way for future exploration in a controlled laboratory set-up. 

\end{abstract}

\maketitle

\section{Introduction}

Wave propagation in inhomogeneous media is a vast and fascinating subject, in which even everyday systems give rise to a variety of beautiful phenomena such as coronae, rainbows and glories. These effects are associated with critical points and caustics at which the ray-optics treatments of wave propagation break down~\cite{FORD1959259, nussenzveig2006diffraction}.
In this paper, we consider surface water waves propagating over a fluid with a varying depth, a problem common in oceanography and coastal engineering, and its associated critical phenomena.

In Ref.~\cite{Berry_tsunami}, Berry considered the propagation of water waves over smooth obstacles, showing that a submerged island acts as a lens for surface waves. The amplification from lensing multiplies the devastating power of tsunami waves, and this amplification is greatest near the focal point at the cusp of the caustic. 
Shallow islands create focal points that are located relatively far from the submerged island. In this region, analytical solutions for the wave profile near the focus point may be found by means of the paraxial approximation \cite{Berry_tsunami}, where all rays contributing to the focusing are assumed to propagate parallel to one another. 

In this work, we move beyond the case of shallow submerged islands to consider more substantial obstacles with a typical relative height $h_0/h_\infty \approx 1/2$, and focal points that are close to, or even on top of, the submerged island. The paraxial approximation is invalid in this case, and we turn instead to methods including geometric optics, numerical simulations, and the Gaussian beam approximation.

\begin{figure}[!h]
 \includegraphics[scale = 0.5]{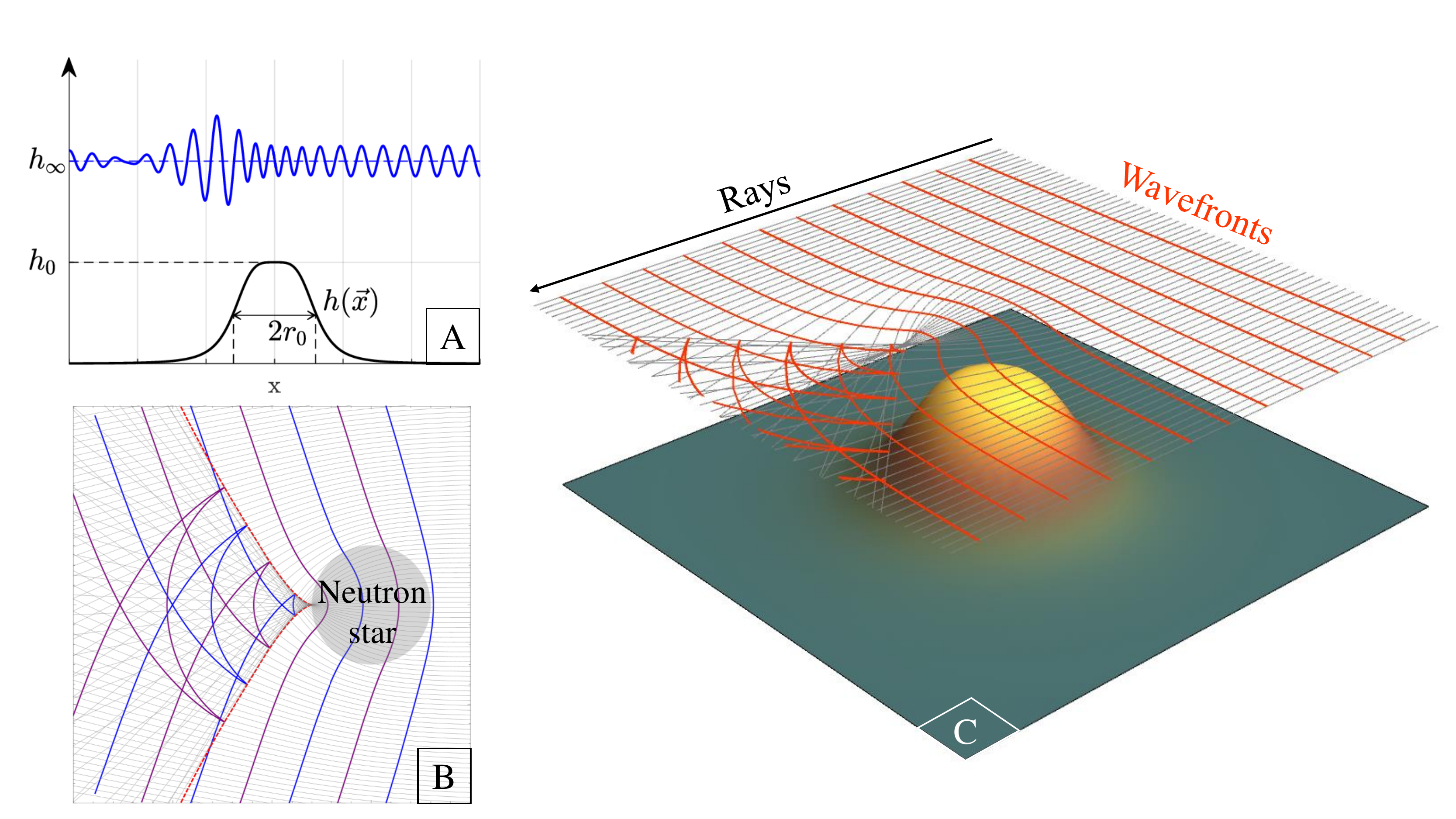}
 \caption{Qualitative description of wave scattering with a focusing object. Panel A depicts a schematic of the underwater island along the axis $y=0$. The blue curves represent the profile of the waves as passing over the submerged obstacle. The wavelength shortens as the water depth decreases; this is followed by an increase in amplitude. Panel B, reproduced from~\cite{Stratton:2019deq}, shows rays and waves in a neutron star spacetime. The blue and purple lines depict the geometrical wavefronts which are retarded by the strong gravitational field of the neutron star. The red dashed curve is the caustic traced out by the cusps in the wavefront. Panel C shows the behavior of rays (in black) and geometrical wavefronts (in red) propagating over the submerged island. Note the striking similarities between panel B and C.}
 \label{fig:schematic}
\end{figure}

The key features of the process are illustrated in Fig.~\ref{fig:schematic}. A wave approaches the island; as the water depth decreases, its speed of propagation decreases also, causing focussing and an associated increase in the wave amplitude  (Fig.~\ref{fig:schematic}A). In passing over the island, the wavefront slows down until, at the focal point, it develops a pair of cusps. These cusps in the wavefront move outwards along a caustic wedge (red dashed line in Fig.~\ref{fig:schematic}B). Inside the caustic wedge, the wave now has leading and secondary fronts. At the focal point, there is substantial amplification; just inside the caustic wedge we anticipate the amplification and diffraction effects commonly associated with rainbows \cite{nussenzveig2006diffraction}.

Remarkably, the focussing of water waves by a submerged island is qualitatively similar, in many respects, to the focussing of gravitational waves by an astrophysical compact object, such as a star or neutron star \cite{Dolan:2017rtj,Stratton:2019deq}. The gravitational potential well of a compact object causes gravitational time dilation, which slows down the wavefront, and consequently generates a focal point and cusp caustic. For our Sun, this focal point is circa 550 a.u.~(astronomical units) distant from its centre. On the other hand, for a neutron star the focal point is expected to be either inside the star, or close to its crust \cite{Dolan:2017rtj,Stratton:2019deq}. In this picture, one may think of the Sun as analogous to the shallow submerged island considered by Berry \cite{Berry_tsunami}, and a neutron star as analogous to the more substantial obstacles considered here.

This analogy extends further. Under certain physical assumptions, water waves propagating in a background fluid flow are described by the equation for a massless scalar field on an effective curved space-time. This observation is the basis of the field of analogue gravity~\cite{Unruh:1980cg,Barcelo:2005fc}, which offers laboratory platforms to experimentally investigate fundamental phenomena occurring in curved space-times, such as Hawking radiation~\cite{Weinfurtner:2010nu,PhysRevLett.117.121301,MunozdeNova:2018fxv}, superradiance~\cite{Torres:2016iee} or cosmological particle production~\cite{PhysRevLett.123.180502,Banik:2021xjn}. 
Here, we show that an underwater island will result in an effective space-time similar to the one of a dense astrophysical object, such as a neutron star.
Moreover, the analogy offers a set of tools that can be applied to both the hydrodynamical system as well the gravitational one. In addition, the analogy offers the possibility to experimentally investigate processes occurring around compact bodies and to test mathematical methods employed to described them.

The paper is structured as follow. In Section~\ref{sec2}, we derive the governing equation for linear shallow water waves propagating over an submerged obstacle and show that it can be seen as the wave equation for massless scalar fields on an effective curved spacetime. We then study the geodesics of this analogue space-time and show that our system exhibits rainbow scattering and focusing at the origin of the cusp caustic. In Section~\ref{sec:3}, we investigate the propagation of waves in our system, both numerically and through semi-analytical methods. In particular, we explore the high-frequency regime using the eikonal (or geometrical) approximation as well as the Gaussian beam approximation which allows us to evolve our geometrical wave through the caustic.
In order to pave the way for experimental investigations of the effects described, we consider the impact of dispersive effects by means of the geometrical ray approximation. This is done in Section~\ref{sec:dispersive_effect}. Finally, Section~\ref{sec:conclusion} concludes our work and discusses possible experimental realisation of our systems. Two appendices contains further technical details on the analytical solution in the case of a parabolic island and on the born approximation to estimate the scattering cross-section.

\section{Wave scattering: methods}\label{sec2}

\subsection{Governing equations}
We consider a stationary, irrotational and inviscid fluid, described by a scalar velocity potential $\phi$ such that the velocity of the fluid is given by $\vec{v} = \vec{\nabla} \phi$.
The surface of the fluid is a 2-dimensional surface with coordinates $\vec{x} = (x,y)$.
An obstacle is placed under the surface such that the fluid depth varies in space. The obstacle is assumed to have a shallow gradient such that derivatives of the fluid depth can be neglected.
Small deformations, $\delta h$, of the fluid interface are related to perturbations of the velocity potential $\phi(\vec{x})$ via $\partial_t \cosh(ih(\vec{\nabla}_{2D})) \phi = -g\delta h$. 
Here $\vec{\nabla}_{2D}$ is the two-dimensional gradient $\vec{\nabla}_{2D} =  \vec{e}_x \partial_x + \vec{e}_y \partial_y$ (we will omit the subscript in the following).
Under these assumptions and neglecting non-linearities, the velocity potential obeys the following wave equation~\cite{Milewski}:
\begin{equation}\label{dispersive_wave_eq}
    -\partial_t^2 \phi + ig \vec{\nabla}\tanh(-i h(\vec{x}) \vec{\nabla}) \phi = 0.
\end{equation}
When considering shallow surface waves, i.e., wavelengths significantly longer than the fluid depth, this reduces to the non-dispersive wave equation
\begin{equation}\label{non_dispersive_wave_eq}
   - \partial_t^2 \phi + c^2(x) \Delta \phi = 0,
\end{equation}
where $\Delta \equiv \vec{\nabla} \cdot \vec{\nabla}$, and $c(x)$ is the propagation speed of the wave given by
\begin{equation}\label{speed}
    c(x) = \sqrt{g h(x)}.
\end{equation}
In the following, we will keep the discussion as general as possible by not specifying a particular profile for the underwater island. Later, to obtain and visualize quantitative results, we will specify the water depth as
\begin{equation}\label{obstacle}
    h(r) = h_\infty - \frac{h_\infty-h_0}{(r/r_0)^n + 1},
\end{equation}
with the default values $h_\infty = 0.04~m$, $h_0 = 0.02~m$, $r_0 = 0.3~m$ and $n=4$. These parameters represent an underwater island which can be set up in existing water tanks~\cite{Torres:2016iee}. This particular class of profile also allows for application of specific analytical techniques, such as the Born approximation. In Appendix~\ref{App:parabola} we present a full analytical calculation for a parabolic island,
\begin{equation}
h(r) = \left\{
                \begin{array}{ll}
                  h_\infty \quad \text{for}\ r\geq R\\
                  h_0 - Br^2 \quad \text{for}\ r\leq R
                \end{array}
              \right. ,
\end{equation}
with $B =(h_0 - h_\infty)/R^2$.

\subsection{Rays and the eikonal approximation\label{sec:eikonal}}
For a general submerged island profile, the wave equation~\eqref{non_dispersive_wave_eq} is in general not solvable in closed form. Nevertheless, approximation methods can be applied to gain insight and to make predictions. A simple but effective method is the eikonal approximation, in which the wave is described in terms of a coherent collection of rays~\cite{Synge63}. As well as being applicable in fluid-mechanical systems, the eikonal approximation has also been successfully applied in analogue-gravity settings in order to describe in wave-vortex interaction scenario~\cite{Torres:2017vaz,Torres:2020ckk} and to describe light-ring mode emission in such systems~\cite{Torres:2019sbr,Torres:2020tzs}. 

The eikonal approximation rests on the assumption that the phase of the wave varies rapidly in comparison with its amplitude. One seeks solutions to the wave equation Eq.~\eqref{non_dispersive_wave_eq} of the form
\begin{equation}
    \phi = A(t,\vec{x})e^{i S(t,\vec{x})/\epsilon},
\end{equation}
where $S$ and $A$ are the local phase and amplitude of the wave, and $\epsilon$ is an order-counting parameter. Expanding the wave equation in a hierarchical fashion in powers of $\epsilon^{-1}$ yields a system of equations for the local phase and amplitude as asymptotic series in $\epsilon$. 
The leading order term in this expansion is the \emph{eikonal equation},
\begin{equation}
 \left(\partial_t S_0\right)^2 - c^2 \left(\nabla S_0\right)^2 = 0, \label{eq:eikonal}
\end{equation}
that determines the leading term in the expansion of the phase, $S_0$. The lines of constant phase $S_0$ are the \emph{wavefronts}. 

The eikonal equation has a natural interpretation as a Hamilton-Jacobi equation, corresponding to the Hamiltonian $H(\vec{x},t; \vec{k},\omega)$ that determines the \emph{rays} of the system (see \cite{Synge63,Torres:2021wud} for detailed discussions). The Hamiltonian is obtained by making the substitution $(\partial_t S_0,\nabla S_0) \rightarrow (-\omega, \vec{k})$ in Eq.~(\ref{eq:eikonal}), to obtain
\begin{equation}
    H = \frac{1}{2} \left( \omega^2 - c(x)^2 k^2 \right) .
\end{equation}
The condition $H=0$ is nothing other than the dispersion relation.

The rays are parametrized curves $(\vec{x}(\tau),t(\tau))$ that are the solutions of Hamilton's equations. In the scattering scenario, we consider a congruence of rays that impinge from infinity, $x\rightarrow\infty$, which are asymptotically parallel and in phase, such that the incoming wavefronts are straight lines orthogonal to the rays. 
Practically, we solve the Hamilton's equation in Cartesian coordinates to find the trajectories of the rays $\vec{x}(\tau)$ as well as the variation of the momentum $\vec{k}(\vec{x}(\tau))$ along the rays. In a time-independent system the associated frequency $\omega$ is constant. Explicitly, we solve the following system of equations
\begin{eqnarray}
\dot{x} = \frac{\partial H}{ \partial k_x}, &\quad& \dot{y} = \frac{\partial H}{ \partial k_y} \\
\dot{k}_x = -\frac{\partial H}{ \partial x}, &\quad& \dot{k}_y = -\frac{\partial H}{ \partial y},
\end{eqnarray}
where the dot represents a derivative with respect to the ray parameter $\tau$ (i.e.~$\dot{x} = dx/d\tau$, etc.). Once the ray and its associated momentum have been computed numerically, we can reconstruct the eikonal phase along the trajectory. This is done using the definition of the momentum $\vec{\nabla} S_0 = \vec{k}$. The eikonal wavefronts are then found as constant phase lines across a congruence of rays.

\begin{figure}[!h]
 \includegraphics[trim=0 0 0 0cm]{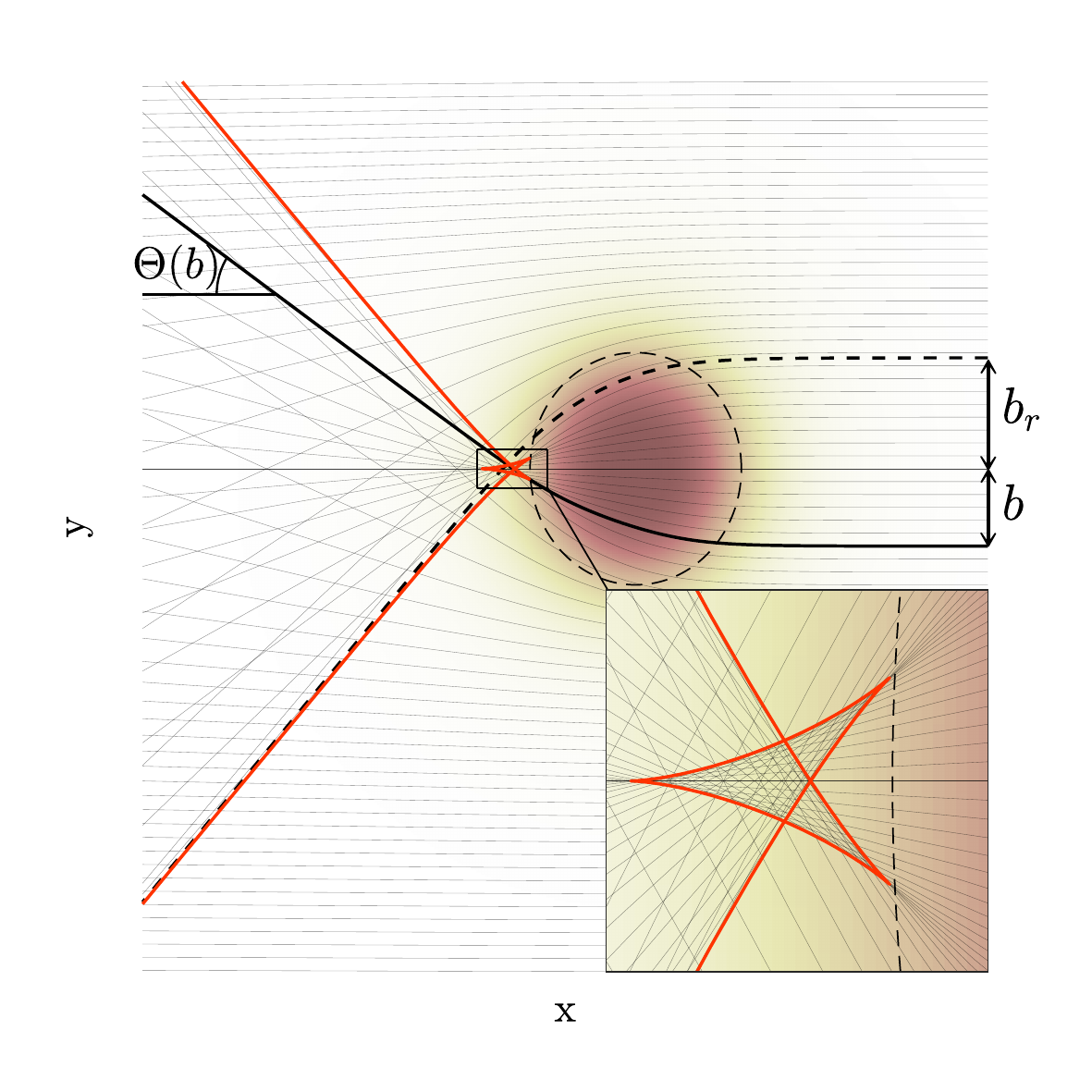}
 \caption{Rays passing over an island form a caustic. A congruence of rays (grey curves) approaches from the right. The solid black line corresponds to a ray passing inside the rainbow wedge, and its associated impact parameter $b$ and deflection angle $\Theta(b)$. The dashed black line shows the rainbow ray, with impact parameter $b_r$, which is maximally deflected (see also Fig.~\ref{fig:deflection}). The red curve highlights the \emph{caustic} where neighbouring rays cross; this is located using the Raychauduri equation (\ref{Raychaudhuri_in_u}). The inset panel reveals the structure of the butterfly caustic.}
 \label{fig:nondispersive_congruence}
\end{figure}

In the case of an axisymmetric submerged island, it is convenient to work with cylindrical coordinates $(r,\theta)$, in which the Hamiltonian becomes
\begin{equation}\label{ham}
    \mathcal{H} = \frac{1}{2}\left(\omega^2 - c^2(r)\left(k_r^2 + \frac{m^2}{r^2} \right)\right),
\end{equation}
where $(\omega,m) = (\frac{dt}{d\tau},\frac{r^2}{c^2(r)}\frac{d\theta}{d\tau} )$ represent the frequency and azimuthal number which are constants of motion, and $k_r$ is conjugate to $r$.
Figure \ref{fig:nondispersive_congruence} depicts a congruence of rays incident on the submerged island from right infinity. 

\subsubsection{The amplitude and the caustic}
To find the leading-order term in the amplitude of the wave, $A_0$, one extends the hierarchical expansions to sub-leading order in $\epsilon$. For a stationary system, of interest here, this equation is
\begin{equation}
    \nabla \cdot (A_0^2 \nabla S_0 ) = 0.  \label{eq:flux}
\end{equation}
This equation implies that the flux of wave action $A_0^2 \nabla S_0$ is conserved along a tube of rays. Using the Hamilton-Jacobi equation and the stationarity condition, the amplitude along a ray is 
\begin{equation}\label{eikonal_amplitude}
    A_0(\tau) \propto \sqrt{\frac{c(\vec{x}(\tau))}{d(\vec{x}(\tau))}},
\end{equation}
where $d(\vec{x}(\tau))$ is the cross-section of the tube of rays.
The cross section of the tube vanishes at points where neighbouring rays meet; here the leading-order amplitude $A_0$ diverges, and the asymptotic expansion breaks down. The \emph{caustic} is the set of all such points.

\subsubsection{Deflection, geometrical scattering and the rainbow angle}
\begin{figure}[!h]
 \includegraphics{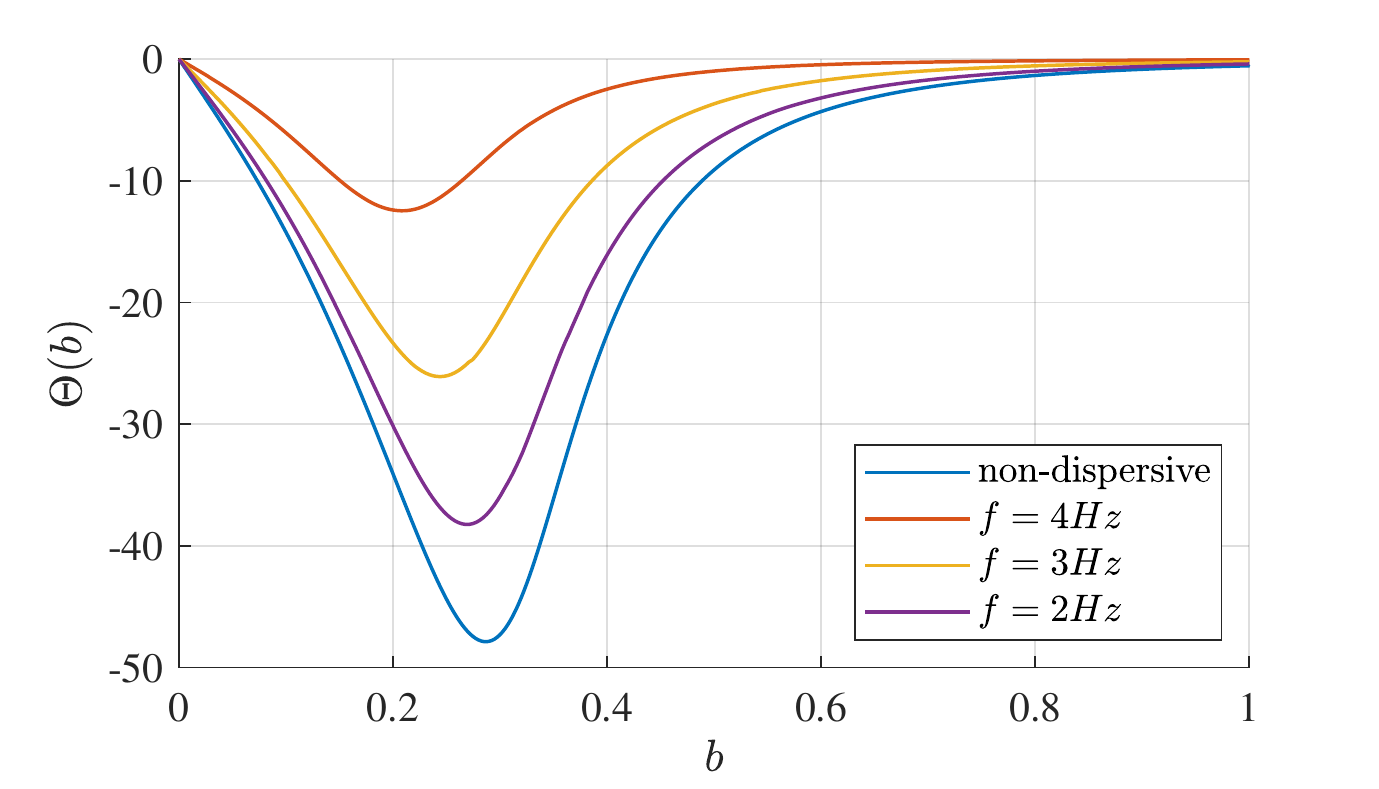}
 \caption{Deflection angle $\Theta$ as a function of the impact parameter $b$. The blue curve corresponds to the non-dispersive case, and is given by Eq.~\eqref{deflection_angle}. The purple, yellow, and red curves correspond to the deflection angle in the dispersive case at $2$, $3$ and $4$ Hz respectively (see~\ref{sec:dispersive_effect} for a discussion of dispersive effects). In every case the deflection angle has an extremum at $b=b_r$, defining the rainbow angle $\theta_r = |\Theta(b_r)|$. Note that dispersion leads to smaller deflection.}
 \label{fig:deflection}
\end{figure}

The ray trajectories provide insight into the scattering of waves with the submerged island. 
Of particular interest is the deflection angle, $\Theta(b)$, for an incident ray with impact parameter $b=mc_\infty/\omega$ (see Fig.~\ref{fig:nondispersive_congruence}).
The deflection angle is obtained by integrating $d\theta/dr$ using Hamilton's equation to get
\begin{equation}\label{deflection_angle}
   \Theta(b) = \pi - 2\int_{r_0}^{\infty} \frac{d\theta}{dr}dr= \pi - 2\int_{r_0}^{\infty}\sqrt{\frac{m^2c^2(r)}{r^4\omega^2 - m^2r^2c^2(r)}} dr,
\end{equation}
where $r_0$ is a turning point satisfying
\begin{equation}
    \frac{c^2(r_0)}{r_0^2}=\frac{\omega^2}{m^2}.
\end{equation}
The deflection angle $\Theta(b)$ is shown in Fig.~\ref{fig:deflection}. 

The \emph{geometrical scattering length} is defined as the density of rays passing into the unit angle $d\theta$. Rays with impact parameter $b$ are deflected by an angle $\Theta(b)$ while the rays with impact parameter $b+db$ are deflected by an angle $\Theta(b) + (d\Theta/db)db$. Consequently, the geometrical scattering length is
\begin{equation}\label{geo_scat_length}
    \left(\frac{d\sigma}{d\theta}\right)_{geo} = \left(\frac{d\Theta}{db}\right)^{-1}.
\end{equation}

As shown in Fig.~\ref{fig:deflection} and Fig.~\ref{fig:nondispersive_congruence}, there is a ray with impact parameter $b_r$ that is maximally deflected, such that $\Theta'(b_r) = 0$ (and $\Theta''(b_r) > 0$). This is known as a rainbow ray \cite{nussenzveig2006diffraction}. This ray determines the rainbow angle of the caustic wedge, $\theta_r = |\Theta(b_r)|$. Formally, the geometrical scattering length diverges at this angle, indicating the breakdown of the method, as expected at any caustic feature.

\subsection{The effective spacetime and the Raychauduri equation\label{sec:spacetime}}

\subsubsection{The effective spacetime}
In accordance with the analogue-gravity paradigm, Eq.~\eqref{non_dispersive_wave_eq} can be rewritten as a Klein-Gordon equation for a scalar field propagating on a curved spacetime as \cite{Unruh:1980cg},
\begin{equation}\label{analogue_grav_eq}
    \frac{1}{\sqrt{-g}} \, \partial_\mu\left(\sqrt{-g} \, g^{\mu\nu}\partial_\nu\phi\right) = 0.
\end{equation}
Here $x^\mu$ are spacetime coordinates (with $x^0 = t$ and $x^i = (\vec{x})^i$), and $g$ is the determinant of the covariant metric tensor $g_{\mu\nu}$ whose inverse is given in contravariant form by
\begin{equation}\label{inverse_metric}
    g^{\mu\nu} = \text{diag}\left(-1,c^2(\vec{x}),c^2(\vec{x})\right).
\end{equation}
In Eq.~\eqref{analogue_grav_eq}, we have used the Einstein summation convention which implies that  repeated indices are summed over. Indices are lowered with $g_{\mu \nu}$ and raised with $g^{\mu \nu}$.  
It is important to note that Eq.~\eqref{analogue_grav_eq} is nothing more than a formal rewriting of the wave equation~\eqref{non_dispersive_wave_eq}. The effective spacetime description is a tool that offers a new point of view on the problem of wave propagation in inhomogeneous media.

\subsubsection{Geometric optics and the null geodesics}
In a similar fashion to Sec.~\ref{sec:eikonal}, one may now seek an asymptotic solution to Eq.~(\ref{analogue_grav_eq}) in the short-wavelength regime. Inserting the ansatz 
\begin{equation}
    \phi = A(x) \exp(i S(x) / \epsilon) 
\end{equation}
into (\ref{analogue_grav_eq}) yields at leading order the eikonal equation
\begin{equation}
    g^{\mu \nu} k_\mu k_\nu = 0 , \label{eq:eikonal2}
\end{equation}
 where $k_\mu \equiv \nabla_\mu S_0$, and $\nabla_\mu$ denotes the covariant derivative on the effective spacetime. By taking a derivative of the eikonal, one obtains the \emph{geodesic equation}
\begin{equation}
    k^\mu \nabla_\mu k^\nu = 0 .
\end{equation}
The integral curves $x^\mu(\lambda)$ that satisfy $d x^\mu / d \lambda = k^\mu$ are \emph{geodesics} of the effective spacetime; and these geodesics are \emph{null} by virtue of (\ref{eq:eikonal2}). In summary, the rays of the eikonal of the physical system of Sec.~\ref{sec:eikonal} correspond with the null geodesics in the effective spacetime. 

The geodesics may be found from the Hamiltonian $H(x^\mu, k_\mu) = \frac{1}{2} g^{\mu \nu} k_\mu k_\nu$, which derives from interpreting the eikonal equation (\ref{eq:eikonal2}) as the corresponding Hamilton-Jacobi equation.
By extending the expansion to sub-leading order, one finds that the amplitude $A_0$ is governed by a transport equation $k^\mu \nabla_\mu A_0 = -\frac{1}{2} \vartheta A_0$, where $\vartheta \equiv \nabla_\mu k^\mu = \Box S_0$ is the \emph{expansion scalar}. 

\subsubsection{Focusing and the Raychaudhuri's equation}
We shall now show that the effective spacetime description yields a practical method for computing the caustic, that is, the set of points where neighbouring rays meet. This is done by solving a transport equation for the expansion scalar $\vartheta$ 
associated with a bundle (or congruence) of geodesics, that quantifies the way neighbouring geodesics converge (or diverge). This transport equation is known as Raychaudhuri's equation.  Following~\cite{Hawking:1973uf}, the Raychaudhuri equation for this system is
\begin{equation}\label{Analogue_Raychaudhuri}
    \frac{d\vartheta}{d\lambda} = -\vartheta^{2} - R_{\mu \nu} k^{\mu} k^{\nu},
\end{equation}
where $k^\mu$ is tangent to a null geodesic, and $R_{\mu \nu}$ is the Ricci tensor, a tensor which describes the local curvature of the effective spacetime (see Appendix~\ref{app:Ricci} for further details). For an arbitrary submerged island, the Ricci tensor is given (see Appendix~\ref{app:Ricci}) in Cartesian coordinates by
\begin{equation}\label{ricci_tensor_matrix}
    R_{\mu\nu} = \begin{pmatrix}
    0 &  0 & 0\\
    0 & \gamma & 0 \\
    0 & 0 & \gamma 
    \end{pmatrix},
\end{equation}
with $\gamma = \frac{1}{2} \left( \frac{\Delta h}{h} - \frac{\nabla h \cdot \nabla h}{h^2} \right)$, where $\Delta h = (\partial_x^2 + \partial_y^2) h$ and $\nabla h \cdot \nabla h = (\partial_x h)^2 +  (\partial_y h)^2$. 
In the case of an axisymmetric island, one can express the coefficient of the Ricci tensor in polar coordinates as $\gamma = 2 f + r \partial_r f$, with $f = \frac{1}{2h(r)r}\frac{dh}{dr}$.

To find the caustic from the Raychauduri equation in practice, we seek points along geodesics at which $\vartheta \rightarrow -\infty$. After the change of variables $\vartheta = \frac{u'}{u}$, where $u = u(\lambda)$ is some function and $u' \equiv \frac{du}{d\lambda}$, finding the caustic reduces to finding points at which $u$ is zero (providing that its derivative is well behaved). The Raychaudhuri equation becomes
\begin{equation}\label{Raychaudhuri_in_u}
    \frac{d^2 u}{d\lambda^2} +  R_{\mu \nu} k^{\mu} k^{\nu} \, u = 0 .
\end{equation}

\subsection{Characterising the caustic\label{sec:caustic}}

The caustic separates the $(x,y)$-plane in two regions with distinct wave profiles (see Fig.~\ref{fig:nondispersive_congruence}). In the region outside of the caustic, each point is connected to the initial wavefront by a single ray, and the phase function is single-valued. Conversely, in the region inside the caustic, each point is connected to the initial wavefront by more than one ray, which results in a multi-valued phase function, and constructive/destructive interference effects.

In the case of interest here, the caustic belongs to the \emph{butterfly caustic} class. Inside the primary region of the butterfly caustic, that is between the rainbow rays, but outside of the star-shaped region, each point is reach by 3 different rays; while inside the star-shape region itself (Fig.~\ref{fig:nondispersive_congruence}, inset), points may be reached by at most 5 different rays~\cite{BERRY1980257}. 
The multivaluedness of the phase means that the geometrical wavefronts are not smooth inside the caustic; rather, they develop cusps and self-intersection as they move through the caustic.

\begin{figure}[!h]
 \includegraphics[trim=0 0 0 0cm]{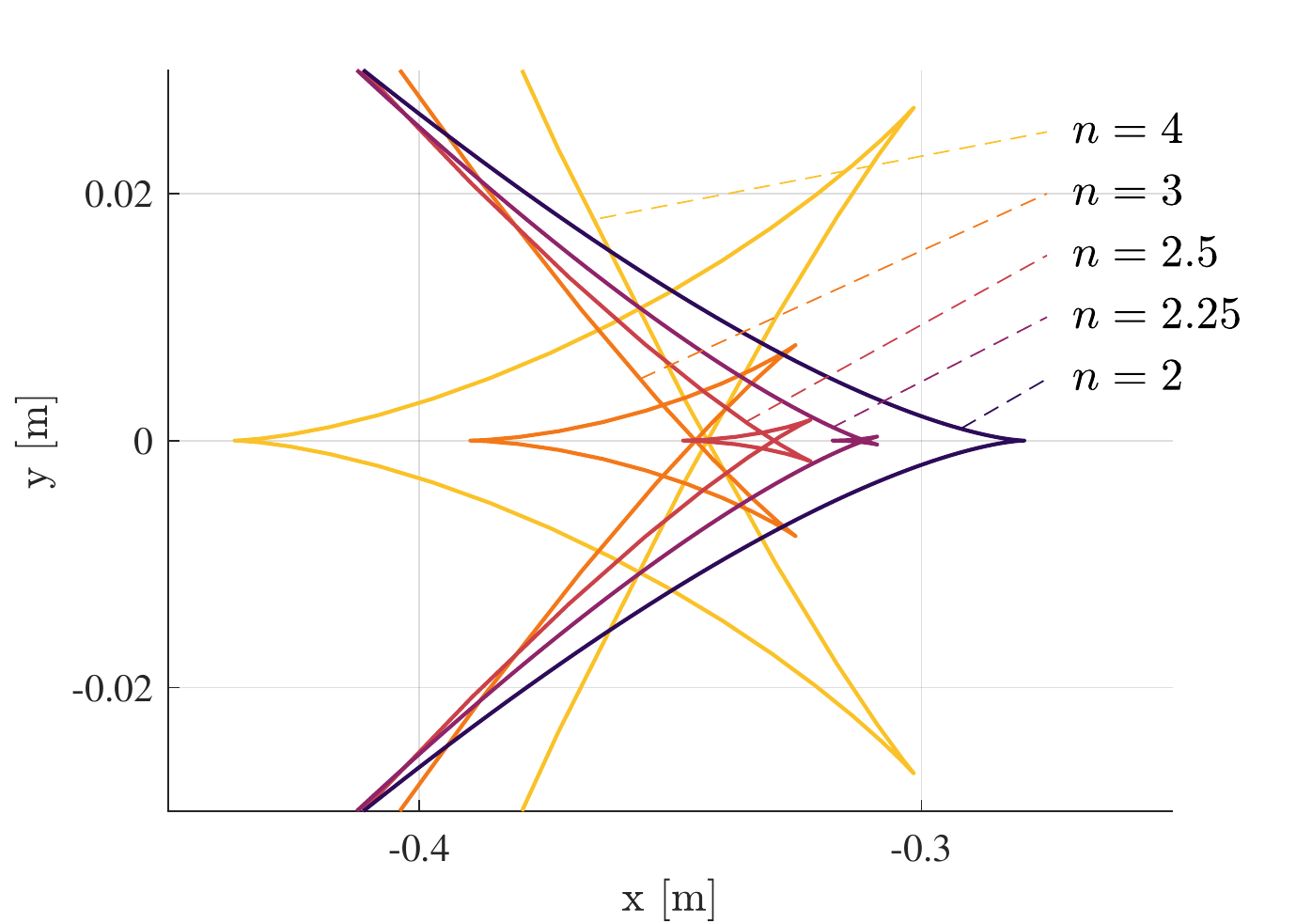}
 \caption{Caustic structure for several island profiles. The profile of the island in Eq.~\eqref{obstacle} is determined principally by the parameter $n$. The caustic morphs from a single cusp to a butterfly as $n$ increases from $2$ to $4$. Here the other parameters are fixed: $h_{\inf} = 0.04$ $\text{m}$, $h_0=0.02$ $\text{m}$ and $r_0 = 0.3$ $\text{m}$. This figure is essentially a view from the top of Fig.~\ref{fig:chrysalis_3d}}
 \label{fig:chrysalis}
\end{figure}

Figure \ref{fig:chrysalis} shows the caustic structure for several different island profiles, in the family of Eq.~\ref{obstacle}. Changing the parameter $n$ changes the shape of the caustic. In particular, the caustic changes from a single cusp to a butterfly as $n$ increases. We now seek to understand this phenomenon at a deeper level.

\begin{figure}[!h]
\includegraphics[trim=0 0 0 0cm]{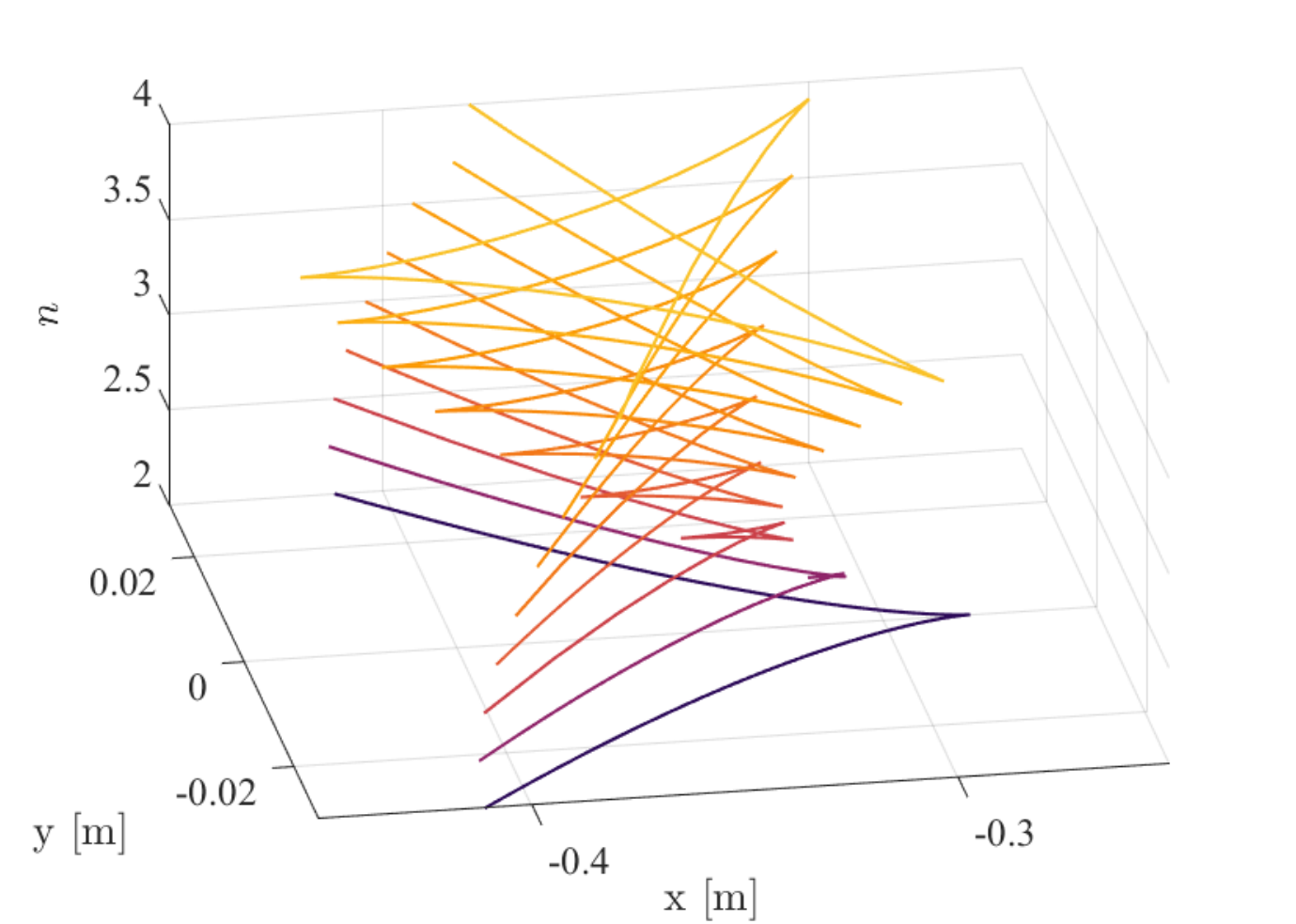}
\caption{Two dimensional surface of the butterfly caustic in the three dimensional $(n,x,y)$-space. The surface is represented as slices for different values of $n$. Each slice results in a curve, depicted in colors ranging from purple to yellow, in the $(x,y)$ plane corresponding to the surface of the water.}
\label{fig:chrysalis_3d}
\end{figure}

The butterfly caustic observed in our system can be understood as a projection of an hypersurface in a four-dimensional space of control parameters onto the two-dimensional $(x,y)$ plane of the water surface. The particular symmetry of the underwater island fixes one of the four parameters ($C_3 = 0$, see below). The caustic is then found by taking a two-dimensional slice through a three-dimensional space. This can be explored by varying the island profile. In particular, by varying the parameter $n$ in Eq.~(\ref{obstacle}), governing the fall off of the island, we shift the height of the slice through the surface, revealing the transition from a cusp caustic to the characteristic star shape of the butterfly caustic. 
Figure \ref{fig:chrysalis_3d} illustrates this point by showing the caustic shape as a function of $n$, which generates a 2D surface in a 3D space.

In catastrophe theory, caustics are found from a generating function, $\phi(s;C)$, where $s = (s_i)$ are state variables and $C = (C_i)$ are control parameters. Essentially, the variables $(s_i)$ parametrize the rays going from an initial surface to a point characterised by the variables $(C_i)$. Note that the control parameters will include the coordinates of the end point but may also contain other parameters governing the media in which rays propagate.
The caustic is then found by looking for singularities of the gradient map, from $s$-space to $C$-space, defined by the condition
\begin{equation}\label{cata1}
    \frac{\partial\Phi(s;C)}{\partial s_i} = 0.
\end{equation}
An end point parametrized by the control parameters $(C_i)$, is a singularity if the Hessian of $\phi(s;C)$ vanishes, that is
\begin{equation}\label{cata2}
    \det \left[ \frac{\partial^2 \Phi}{\partial s_i \partial s_j}\right] = 0.
\end{equation}
One of the main result of catastrophe theory is that caustics are divided into equivalence classes, and all elements of a class can be deformed into one another smoothly. Each equivalence class is described by a generating function, in the form of a standard polynomial. The butterfly caustic has a single state parameter, $s$, and its standard polynomial is~\cite{BERRY1980257}
\begin{equation}
    \Phi(C_i, s) = s^6 + C_4 \frac{s^4}{4} + C_3 \frac{s^3}{3} + C_2 \frac{s^2}{2} + C_1 s.
\end{equation}
From Eqs.~\eqref{cata1} and~\eqref{cata2}, we can express $C_1$ and $C_2$ as functions of $(s,C_3,C_4)$. For a fixed value of $C_3$, and $C_4$, the caustic is found as a curve parametrised by $s$ in the $(C_1,C_2)$ plane. The transition from the cusp to the butterfly caustic can be seen by fixing $C_3=0$ and varying $C_4$ for some positive to some negative value. For $C_4 > 0$, the caustic in the $(C_1,C_2)$ plane is a cusp caustic, while for $C_4 < 0$ it is a butterfly caustic. The \emph{chrysalis point}, from which the butterfly emerges, is at $C_4 = 0$.

From the previous description we can relate $(C_1,C_2)$ to the Cartesian coordinates $(x,y)$ and $C_4$ to the parameter $n$ governing the island profile. In particular, $C_4$ will be a monotonically decreasing function of $n$, such that $C_4(n<n_{chrysalis})>0$ and $C_4(n>n_{chrysalis})<0$.

\subsection{The Gaussian Beam Approximation}

As we have seen, approximating the amplitude of the wave in the vicinity of caustics is impossible by means of the geometrical approximation, which predicts an infinite amplitude at the focus point.
One can deal with such infinities in some cases by modifying the ray method and introducing special functions. 
This is the case in the Airy treatment of the rainbow scattering for example, which can be applied to a variety of caustic problems in one spatial dimension. 
This method is based on obtaining an approximate wave equation in the vicinity of the caustic.
Another method of curing the singularities of the geometrical description is to reintroduce some ``wave flesh onto the classical bones"~\cite{Berry_1972}. Instead of considering the wave as a congruence of rays, one can consider it as being a collection of \emph{beams}. Each beam is centred on the underlying rays, and the amplitude is not located exactly on the ray but on a Gaussian profile transverse to the ray. This Gaussian beam can spread and focus depending on the inhomogeneities of the media. For this reason, the method is called the \emph{Gaussian beam approximation}~\cite{erven1982ComputationOW,popov1982new}.
More precisely, the wave equation~\eqref{non_dispersive_wave_eq} is reduced to a local wave equation around each ray. 
The essential ingredients of the Gaussian beam approximation are outlined below.

First, define an orthonormal coordinate basis $(\vec{e}_s,\vec{e}_n)$ adapted to each ray, where $\vec{e}_s$ is tangent to the ray and $\vec{e}_n$ is transverse, such that $\vec{e}_s.\vec{e}_n = 0$. A point in the vicinity of the ray is located with coordinate $(s,n)$ where $s$ is the arc-length along the ray and $n$ the displacement along $\vec{e}_n$.
In this coordinate system, we look for solution to the wave amplitude of the Gaussian form
\begin{equation}
    A(s) = \sqrt{\frac{c(s,n=0)}{q(s)}}\exp\left[\frac{i}{2}n^2\Gamma(s)\right],
\end{equation}
where $\Gamma(s) = \frac{p(s)}{q(s)}$ and $(p(s),q(s))$ are unknown functions which obey the following system of differential equations:
\begin{equation}
    \frac{dq}{ds} = c(s,0)p, \: \: \: \: \frac{dp}{ds} = - c(s,0)^{-2} \frac{\partial^2 c}{\partial n^2}(s,0)q .
\end{equation}

Here we are interested in the amplitude of the ray passing through the caustic, which we take to be located on the $y=0$ axis.
For such central ray, the arc-length, $s$, is precisely equal to its displacement from the initial point hence $s = x_{0} - x$. Now, along the central ray, we take the normal to be $\vec{n} = (0,1)$, and therefore the normal coordinates can be substituted directly as $y$, because $\frac{dy}{dn} = 1$. In Cartesian coordinates $(p,q)$ now obeys the following differential equations
\begin{equation}\label{GB_eq}
    -\frac{dq}{dx} = c(x,0)p, \: \: \: \: \frac{dp}{dx} = c(x,0)^{-2} \frac{\partial^2 c}{\partial y^2}(x,0)q.
\end{equation}

In the traditional ray method, $q$, $p$ and $\Gamma$ are real and $q$ vanishes at the caustic points which causes the amplitude to diverge. 
In order to avoid the singularity, $p$ and $q$ must be complex valued, and we can think of the Gaussian beam as a collection of \emph{complex} rays~\cite{deschamps1971gaussian}. 
Since the coefficients in Eq.~\eqref{GB_eq} are real, it follows that the initial condition must be complex.
It turns out that the evolution of the Gaussian is characterised by a single complex parameter, which can be represented via two real numbers. Those two real numbers can be interpreted as the initial effective half-width of the Gaussian beam and the distance between the initial position of the beam and the point where the half-width is minimum (i.e., the distance between the initial point and the caustic)~\cite{erven1982ComputationOW}.
Since we are only interested in the amplification at the caustic, the initial half-width can be normalised and we are left with a single parameter controlling the evolution of the Gaussian beam.
To finalise setting-up the initial condition, we choose the minimum half-width of the Gaussian beam to be located at the focus point determined using the eikonal approximation.

\subsection{Wave scattering by an analogue neutron star\label{sec:partial-wave}}
As shown in Sec.~\ref{sec:spacetime}, surface waves propagating over the submerged island obey a massless scalar field equation in a curved spacetime, Eq.~\eqref{analogue_grav_eq}.
We consider here a monochromatic solution, $\phi = \text{Re} \, e^{i\omega t} \phi_\omega(\vec{x})$, which can be constructed from a sum over partial waves, as follows:
\begin{equation}
    \phi_\omega(\vec{x}) = \sum_{m=-\infty}^{+\infty}\frac{1}{\sqrt{r}}\phi_{\omega,m}(r) e^{im\theta}.
\end{equation}
Here, the radial profiles $\phi_{\omega,m}(r)$ satisfy the following radial equation
\begin{equation}\label{waveeq_mode}
gh(r)\phi_{\omega,m}'' + \left( \omega^2 - gh(r)\frac{m^2 - 1/4}{r^2}\right) \phi_{\omega,m} = 0.
\end{equation}
Far from the obstacle, the depth of the water tends to a constant, $h(r)\rightarrow h_\infty$ and the wave equation~\eqref{waveeq_mode} takes the form
\begin{equation}
    \phi_{\omega,m}'' + c_\infty^2 \phi_{\omega,m} = 0,
\end{equation}
with $c^2_\infty = \omega^2/g h_\infty$, which admits the solutions
\begin{equation}\label{asyminf}
\phi_{\omega,m}(r) \sim A^{\text{out}}_{\omega,m}e^{ik_\infty r} + A^{\text{in}}_{\omega,m}e^{-ik_\infty r}, \quad r\rightarrow\infty,
\end{equation}
with $c_{\infty}k_\infty = \omega$. $A^{\text{out/in}}_{\omega,m}$ represent the (complex) amplitude of radially outgoing/ingoing modes.

We are interested in the scattering of an incident monochromatic plane wave. Hence, we seek solutions which far from the island, are a superposition of a plane wave propagating towards $x\rightarrow - \infty$ and a radially-outgoing scattered component,
\begin{equation}\label{scattered_wave_decomposition}
\phi \sim e^{-i\omega t} \left(e^{ikx} + f_{\omega}(\theta)\frac{e^{ikr}}{\sqrt{r}} \right)
\end{equation}
The function $f_\omega(\theta)$ is the scattering amplitude, which can be expressed as a partial-wave sum, 
\begin{equation}\label{def_scat_amp}
f_{\omega}(\theta) =\left( \frac{1}{2i\pi k}\right)^{1/2} \sum_{m=-\infty}^{+\infty} \left( e^{2i\delta_m} - 1\right) e^{im\theta}.
\end{equation}
The rotational invariance of the island profile implies that the phase shifts are symmetric, $\delta_m = \delta_{-m}$,  and we can rewrite the expansion of the scattering amplitude as
\begin{equation}\label{scattering_cs}
f_{\omega}(\theta) = \sum_{m=0}^{\infty} a_m\cos(m\theta), \quad \text{with} \quad a_m = \left\{
                \begin{array}{ll}
                  \left( \frac{1}{2i\pi k}\right)^{1/2}\left( e^{2i\delta_0} - 1\right) \quad \text{for}\ m=0, \\
                  \left( \frac{2}{i\pi k}\right)^{1/2}\left( e^{2i\delta_m} - 1\right) \quad \text{for}\ m>0.
                \end{array}
              \right.
\end{equation}
By decomposing the incoming plane wave onto the azimuthal basis as, $e^{ikx} = \sum_{m}i^m J_m(kr)e^{im\theta}$, and using the asymptotic form at infinity of the Bessel functions of the first kind, $J_m(kr)$, the phase shifts are found in terms of the ingoing/outgoing amplitudes in Eq.~\eqref{asyminf}, 
\begin{equation}\label{phase_shift_def}
e^{2i\delta_m} = i(-1)^m \frac{A^{\text{out}}_{\omega,m}}{A^{\text{in}}_{\omega,m}}.
\end{equation}
In the case of a submerged obstacle, the phase shifts are real ($\delta_m \in \mathbb{R}$) since there is no absorption or amplification and hence $|A^{\text{out}}/A^{\text{in}}| = 1$.
\newline
\newline
\textbf{Geodesic phase shifts}
\newline
\newline
 As we have just seen, in the high frequency/large $m$ limit, waves can be described as a collection of rays. Hence, we can express in this limit the phase shifts in terms of the properties of the rays. In the large-$m$ regime, there is the following correspondence between the deflection angle~\eqref{deflection_angle} and the phase shifts,
 \begin{equation}
     \Theta(m) = 2\frac{d\delta_m}{dm}.
 \end{equation}
 
 We show in Appendix~\ref{App:Born} that, in the large $m$ limit, and for the general shape of underwater island we consider in Eq.~\eqref{obstacle}, the deflection function behaves as $\Theta = \mathcal{O}(m^{-n})$, where $n$ governs the fall off of the obstacle at infinity. This implies that, for $n\neq 1$, the phase shifts behaves as $\delta_m =\mathcal{O}(m^{n-1})$ and will tend to zero as $m \rightarrow\infty$. Conversely, in the case of gravitational body of mass $M$, the deflection angle falls off as $\sim 4M/b$, which leads to a logarithmic behavior of for the phase shifts, and a Rutherford-like ($\theta^{-4}$) divergence in the scattering cross section in the forward direction. 

\section{Wave scattering: simulation and comparison}\label{sec:3}

\subsection{Numerical Method}\label{sec:numerics}

We solved the 1D wave equation~\eqref{waveeq_mode} with a numerical method to obtain the radial profiles of the partial waves $\phi_{\omega,m}(r)$. We start with appropriate initial conditions at $r= \epsilon\ll 1$, by selecting initial values $\phi_{\omega,m}(\epsilon)$ and $\phi'_{\omega,m}(\epsilon)$ which determined from the solution to the approximate wave equation at small radius,
\begin{equation}
    \phi''_{\omega,m} - \frac{\left(m^2 - 1/4\right)}{r^2} \phi_{\omega,m} = 0.
\end{equation}
This equation admits the regular solution $\phi_{\omega,m} \sim r^{1/2+|m|}$.
We then  integrate numerically Eq.~\eqref{waveeq_mode} using the {\tt NDSolve} solver in Mathematica into the far-field.
The ingoing and outgoing amplitudes $A^{\text{in/out}}_{\omega,m}$ in Eq.~\eqref{asyminf} are obtained by matching the numerical solution onto the generalized series solutions,
\begin{equation}\label{genseriessol}
    \phi^{in} = \left(\phi^{\text{out}}\right)^* =  e^{-ik_\infty r} \sum_{j = 0}^{N} b_j r^{-j} .
\end{equation}
The coefficients $b_j$ are obtained from the recurrence relation that is obtained by substituting the generalized series solution~\eqref{genseriessol} into Eq.~\eqref{waveeq_mode} and expanding order-by-order in powers of $1/r$. 
Finally, we calculate the phase shifts $\delta_m$ by inserting the ingoing and outgoing amplitudes in Eq.~\eqref{phase_shift_def}.

From the set of radial profiles, we reconstruct a monochromatic plane wave incident on the submerged island by matching the relative phases of each mode and requiring that it satisfies Eq.~\eqref{scattered_wave_decomposition} in the far field. Explicitly, the incident plane wave is reconstructed by evaluating the following sum
\begin{equation}
    \phi_\omega(r) = \sum_{m=-\infty}^{\infty} \frac{(-1)^m}{A_{\omega,m}^{in}}\sqrt{\frac{i}{2\pi k _\infty}}\frac{\phi_{\omega,m}(r)}{\sqrt{r}}e^{im\theta}.
\end{equation}
When reconstructing the incident plane wave, as well as the scattering cross-section from the partial wave expansion, we numerically compute the radial profile of a finite number of $m$ modes, up to $m \le M$, and then use Bessel functions, which are the solutions to the wave equation for large $m$, to complete the sum up to second cut-off, $M2$, which is greater than $M$. The value of $M$ is chosen such that the phase shift obtained numerically satisfy $|1 - e^{2i\delta_m}| < 10^{-3}$, and $M_2$ such that the Bessel function are negligible in the region of interest.

\subsection{Results}

\begin{figure}[!h]
 \includegraphics{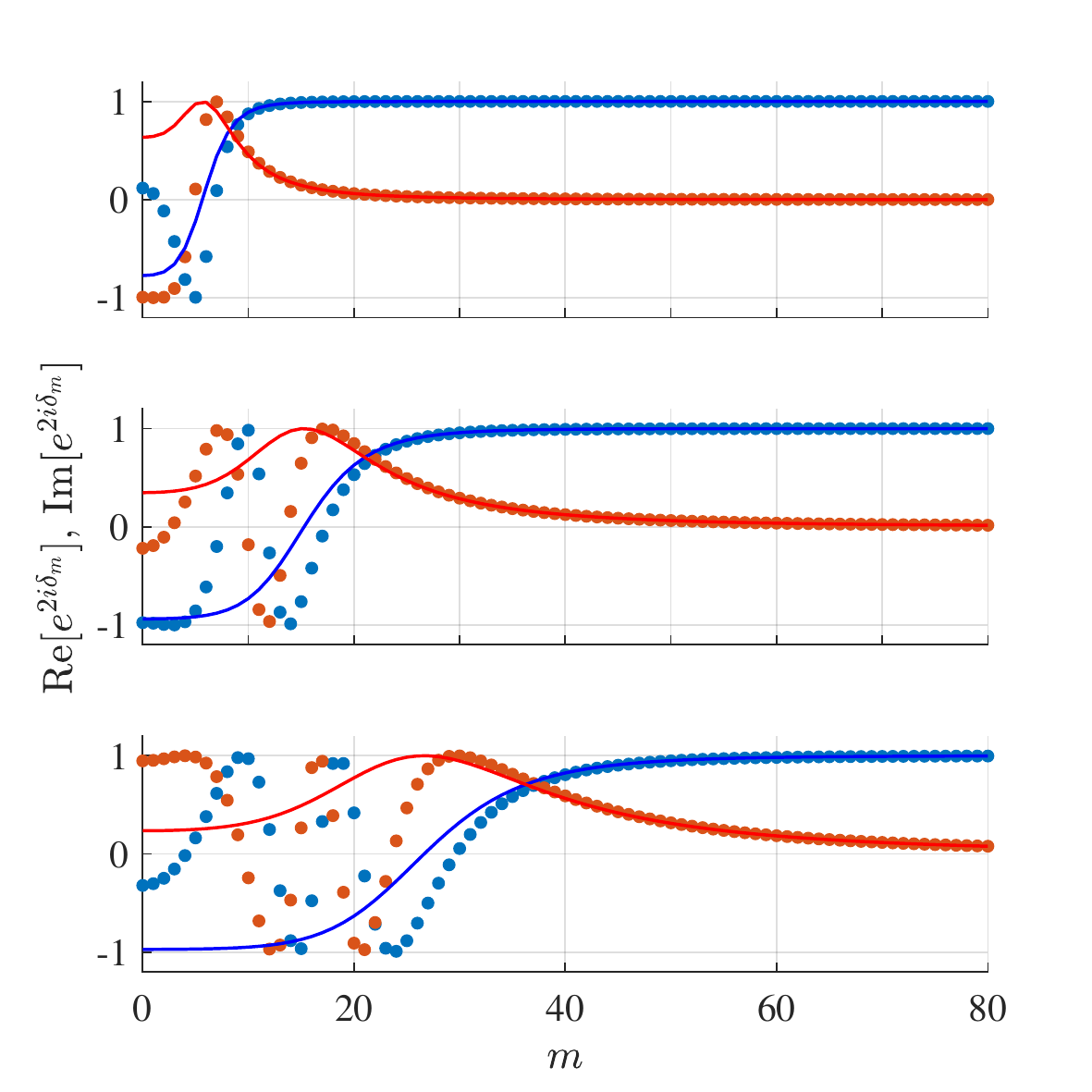}
 \caption{Scattering coefficients $e^{2 i \delta_{m}}$ for the frequencies  $f=2$Hz (upper), $f=4$Hz (middle), and $f=6$Hz (lower). The plots compare the numerically-determined coefficients (dots) with those obtained via the Born approximation (lines) (see Appendix \ref{App:Born}). The real (imaginary) parts are shown in blue (red).}
 \label{fig:phase_shifts}
\end{figure}

Figure \ref{fig:phase_shifts} shows the phase shifts obtained numerically, and the comparison with values from Born approximation in Appendix \ref{App:Born}, for different frequencies: $f=2$Hz, $f=4$Hz, and $f=6$Hz. We can see the qualitative agreement between the approximate estimate and the numerical value of the phase-shifts.

\begin{figure}[!h]
 \includegraphics{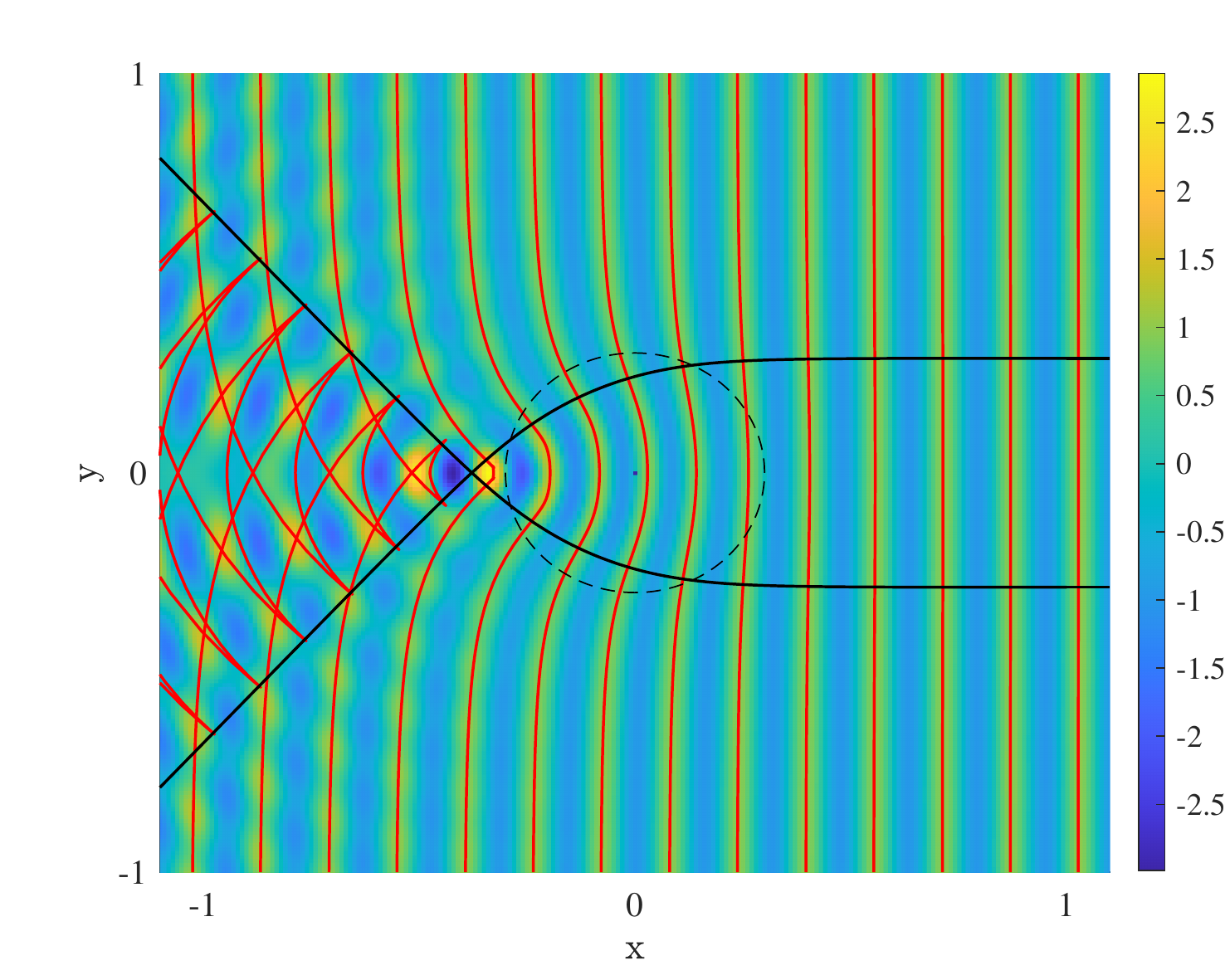}
 \caption{Simulation of a wave passing over a submerged island. The plot shows an incident plane wave with frequency $f=4$Hz. The colours (blue/green) represent the wave amplitude, in arbitrary units, obtained via the partial wave method. The eikonal wavefronts are shown as red lines. The black lines indicate the rainbow rays, and the dashed black circle represents the typical size of the underwater island, $r_0$. There is a good qualitative agreement between the eikonal and the numerical wavefronts.}
 \label{fig:wave_pattern_4Hz}
\end{figure}

Figure \ref{fig:wave_pattern_4Hz} shows an incident plane wave of frequency $f=4$Hz encountering an underwater island. The focusing of the incident plane wave and the resulting increase in amplitude is clearly visible in Fig.~\ref{fig:wave_pattern_4Hz}, and for this frequency the amplification factor is $\sim 3$. Interference effects are visible within the rainbow wedge. The eikonal wavefronts (red lines) closely track the undulations in the full numerical solution. The eikonal wavefronts were found by starting from an initial wavefront on the right hand side of Fig.~\ref{fig:wave_pattern_4Hz} and then ray-tracing using Hamilton's equations. The island parameters used in the partial wave expansion and the eikonal methods are identical and are given in Eq.~\eqref{obstacle}.  

\begin{figure}[!h]
 \includegraphics{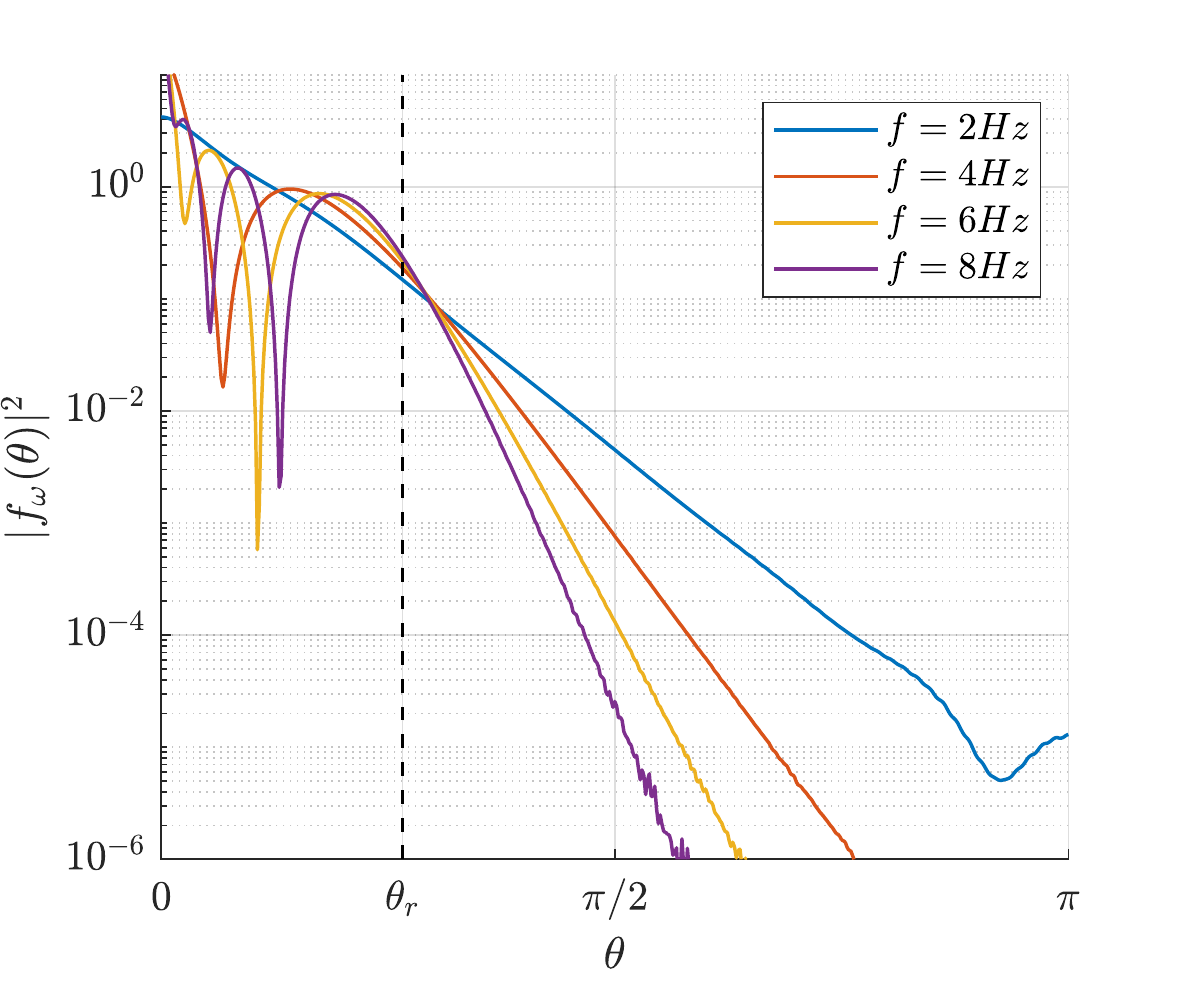}
 \caption{Scattering cross section of an incident plane wave on the submerged island for various frequencies. The partial sum is computed up to $N=300$ with the first $80$ modes solved numerically and the contribution for the higher $m$ is estimated using the Born approximation. The vertical dashed curve represent the location of the rainbow angle $\theta_r$. For angles $\theta < \theta_r$, the scattering cross section has an oscillatory behavior; for $\theta > \theta_r$ it decays exponentially.}
 \label{fig:scat_cross}
\end{figure}

Figure \ref{fig:scat_cross} shows the scattering cross section of an incident plane wave on the submerged obstacle for the frequencies $f=2$Hz,$f=4$Hz, $f=6$Hz, and $f=8$Hz. The rainbow angle $\theta_r$ is indicated by the vertical dashed line. Inside the rainbow angle ($\theta < \theta_r$), the scattering cross section has an oscillatory behaviour, whereas outside the angle it has an exponential fall-off, indicatiing a shadow region. This is the typical behaviour in rainbow scattering~\cite{Dolan:2017rtj,Stratton:2019deq,FORD1959259}. 

The scattering cross section shown in Fig.~\ref{fig:scat_cross} was computed using the partial wave expansion given by Eq.~\eqref{scattering_cs} and summed over $0<m<300$. The phase shift are obtained numerically for $m\leq80$, and via the Born approximation for $m>80$.
The noise visible in the scattering cross section of the high frequency for large angle arises from the discrepancy between the numerical phase-shift and the Born approximation for $m\sim 80$.

\subsection{Amplitude at the caustic}

\begin{figure}[!h]
 \includegraphics{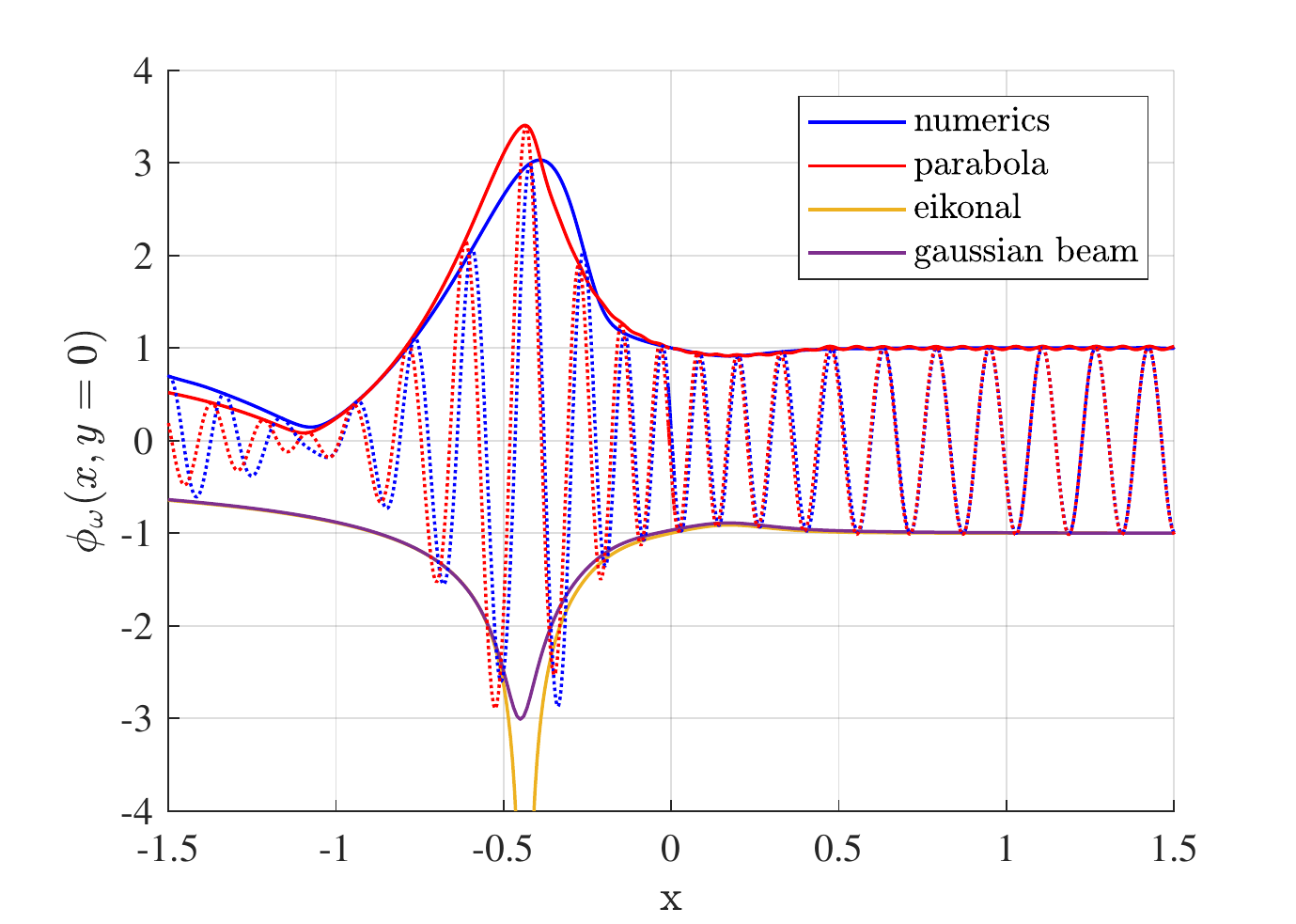}
 \caption{Wave profile propagating to the left evaluated along the $x$-axis and $y=0$ with a frequency $\omega = 8\pi$. The blue curves represent the numerical profile obtained by the partial wave expansion method and the red curves is the wave profile obtained analytically after approximating the underwater island with a finite size parabola (see Appendix.\ref{App:parabola}). The dotted curves represent the real part of the wave profile while the solid curves depicts the absolute value. The purple/yellow envelope depicts the amplitude of the wave in the Gaussian beam/eikonal approximation respectively (the negative sign is only for the clarity of the figure). }
 \label{fig:caustic_amplitude}
\end{figure}

Figure \ref{fig:caustic_amplitude} shows the wave profile along the $y=0$ axis for a wave with angular frequency $\omega = 8\pi$. The blue curve shows the numerical profile obtained by the partial wave expansion method, and the red curve shows the wave profile obtained analytically after approximating the underwater island with a finite size parabola (see Appendix.\ref{App:parabola}). 
The purple/yellow envelope depicts the amplitude of the wave in the Gaussian beam/eikonal approximation, respectively. 
The eikonal approximation of the amplitude diverges at the focus point, whereas the Gaussian beam remains finite and provide an overall good description of the profile of the wave and its amplification. 

The numerical results, as well as the parabolic and Gaussian-beam approximations, reveal three interesting feature of the profile. First, there is a clear increase in amplitude downstream of the obstacle, near the cusp of the caustic. Here there amplification by a factor approximately 3, relative to the original wave amplitude. Second, there is a smaller decrease in amplitude as the waves approach the obstacle. This decrease in amplitude is also present in the eikonal approximation, and can be understood from Eq.~\eqref{eikonal_amplitude}; it is due to the fact that the propagation speed decreases faster than distance between neighbouring rays. Finally, we observe a dip in the amplitude downstream from the caustic cusp. The minimum (near $x=-1$ in Fig.~\ref{fig:caustic_amplitude}) is seen in the numerical simulation, as well as in the parabola approximation. Qualitatively, this dip is due to destructive interference between multiple wavefronts inside the rainbow wedge. It is not present in the Gaussian beam profile, nor in the eikonal profile, because the contribution from secondary rays has not been included.

\section{Dispersive effects}\label{sec:dispersive_effect}
\subsection{From geodesics to rays}

Thus far, we have focused on the case of water waves with a linear dispersion relation, which allowed us to establish a precise analogy between our system and that of gravitational waves propagating through compact bodies.
Of course, this linear dispersion relation is an approximation (one amongst many) to the physical system, and it is well known that surface water waves are subject to dispersive effects.
Considering the impact of dispersion in a general setting is beyond the scope of this paper; however, if we restrict attention to the behaviour of high- frequency waves, then by means of the eikonal approximation, it is possible to extend the analysis of Sec.~\ref{sec:eikonal} to the dispersive regime.

It was shown in Ref.~\cite{Torres:2017vaz} that the trajectories of the  ``particles'' that make up the eikonal waves can be obtained from the dispersive Hamiltonian given by
\begin{equation}
    \mathcal{H}_{D} = \omega^2 - gk\tanh(h(\vec{x})k) = 0,
\end{equation}
with $k = \sqrt{k_{x}^{2}+k_{y}^{2}}$. The eikonal trajectories, which previously corresponded to geodesics of the effective space-time, are now called \emph{rays}, since one cannot define a single effective metric for which all the rays would be the geodesics. We can find the rays, and thus the eikonal wavefronts by solving Hamilton's equations, as before. 

\begin{figure}[!h]
 \includegraphics{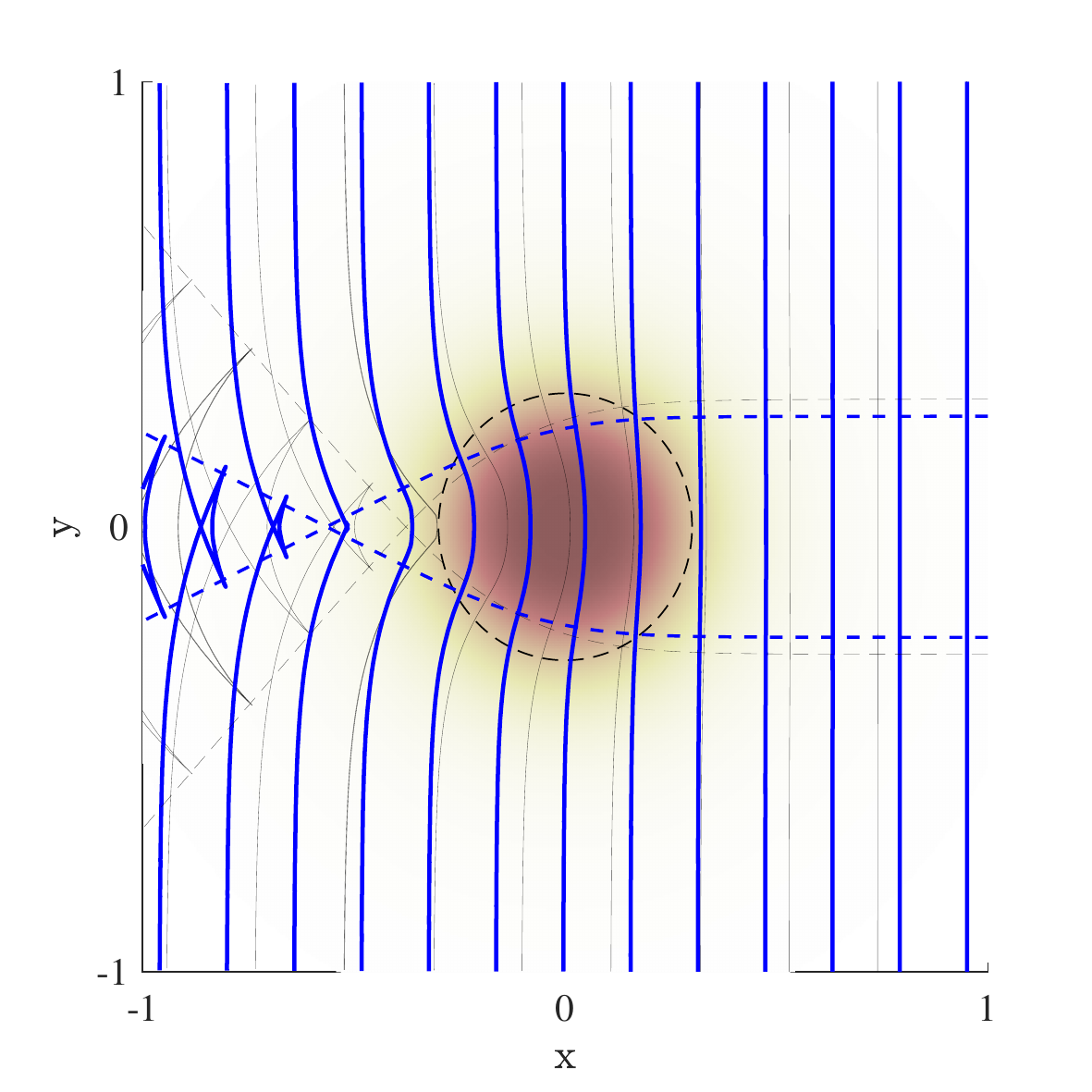}
 \caption{The effect of dispersion on the eikonal wavefronts. The eikonal wavefronts for an incoming wave with frequency $f=3\text{Hz}$ are shown in both  the linear and dispersive regimes, as black and blue solid curves, respectively. The dashed curves represent the rainbow ray in the linear (black) and dispersive (blue) cases. The background colour map and dashed circle indicates the height and shape of the submerged obstacle. Dispersive effects lead to a narrowing of the rainbow wedge, and a cusp focal point that is located further away from the obstacle.}
 \label{fig:disp_vs_nodisp}
\end{figure}

Figure \ref{fig:disp_vs_nodisp} shows the eikonal wavefronts for the dispersive and non-dispersive systems, for an incoming wave of frequency $f=3\text{Hz}$. We may draw two general observations from this figure. First, that the qualitative behaviour between the two regimes is similar, that is, we still observe the presence of a rainbow ray (see also Fig.~\ref{fig:deflection}), and a time delay leading to a focusing of the wave. Second, that the dispersion, or more precisely the subluminal dispersion of water waves, leads to a smaller deflection of the rays by the underwater island. This behaviour can be understood in terms of geometrical optics and Snell's law $n_D \theta_D = n_L \theta_L$ (for small angles), where $\theta_{D,L}$ is the refracted angle and $n_{D,L}$ the refractive index for the dispersive and linear regime respectively. Since the refractive index is inversely proportional to the wave speed, we have that $n_D>n_L$ which implies that $\theta_D<\theta_L$. A consequence is that, in the dispersive case, the rainbow angle is narrower, and the focal cusp is further from the scattering centre.

\section{Conclusion}\label{sec:conclusion}
In the preceding sections, we have analysed the scattering of surface water waves passing over submerged obstacles. As Berry found \cite{Berry_tsunami}, underwater islands act as lenses which focus and amplify incident waves. In the short-wavelength limit, this leads to the formation of caustics near which the power of the wave (or of the tsunami) is enhanced. We have shown here that a submerged island of significant height relative to the water depth will generate strong focusing immediately downstream (see Fig.~\ref{fig:nondispersive_congruence}). Moreover, we find that for substantial islands, the caustics formed can be of cusp or butterfly type, with the latter emerging for steep-sided islands (see Fig.~\ref{fig:chrysalis}). 

We simulated the wave scattering process in the linear regime using a partial-wave decomposition into $m$-modes (Sec.~\ref{sec:partial-wave}). The key features of the scattering patterns are described by several semi-analytical techniques. The eikonal approximation yields wavefronts that are a good match to those seen in the simulation (see Fig.~\ref{fig:wave_pattern_4Hz}). However, the eikonal amplitude diverges at the caustic,  and the eikonal short-wavelength assumption breaks down where neighbouring rays intersect. To circumvent this issue, we adopted the Gaussian beam approximation, which yields a valid prediction for the wave profile across the focal point (the cusp). In Fig.~\ref{fig:caustic_amplitude}, we observed robust qualitative agreement between the Gaussian beam approximation and the numerical simulation. To augment these approximation schemes, we also derived an exact analytical expression for waves propagating over an island of parabolic shape (see Appendix \ref{App:parabola}).

Remarkably, this hydrodynamical system has features in common with strong-gravity systems in astrophysics. In the absence of dispersion and dissipation, the equations governing water waves propagating over a fluid with varying height are mathematically equivalent to those governing a massless scalar field in an effective 2D spacetime (see Sec.~\ref{sec:spacetime}). The effective spacetime is qualitatively similar to (a 2D slice through) the spacetime of a massive, dense body, such as a neutron star. Consequently, the wave scattering patterns possess similar features. For example, in both cases we expect \emph{rainbow scattering}, which is the manifestation of constructive/destructive interference effects on the inside of the rainbow wedge associated with a maximally-deflected ray. We have characterised this effect in our system and observed its presence in our numerical simulation (see Fig.~\ref{fig:scat_cross}). By comparison with the results of Refs.~\cite{Dolan:2017rtj,Stratton:2019deq} in the gravitational context, we find that water waves passing over a submerged island closely resemble gravitational waves focussed by the spacetime curvature of a neutron star; with the most important differences arising from the dimensionality of the systems (2D vs 3D). 

The analogue-gravity description is more than a mathematical curiousity, however. We showed that by solving the Raychaudhuri equation in the effective spacetime, we can locate and characterise the caustic (see Sec.~\ref{sec:caustic}). The Raychaudhuri equation is most familiar in relativistic context in the context of (Penrose-Hawking) singularity theorems. It is a transport equation that describes the rate of change of the cross-sectional area of a congruence of rays. The caustic is the set of points where that cross-sectional area passes through zero.

Finally, we considered the effect of dispersion on the geometrical picture of wave propagation (Sec.~\ref{sec:dispersive_effect}). We established that the key features of the scattering process are modified but not eliminated by dispersive effects. In Fig.~\ref{fig:disp_vs_nodisp}, we observe a similar convergence of rays leading to focusing downstream of the obstacle, and the rainbow scattering is still present; the main difference here is the narrowing of the rainbow angle. Consideration of dispersive effects is an important step towards an experimental realisation of a neutron-star analogue in a wavetank, and direct measurements of the predicted phenomena.

In the results presented in this work, we have selected physical parameters ($h_\infty$, $h_0$ $r_0$, $n$ and $f_0$) that we anticipate are relevant to (future) wavetank experiments. Any implementation in physical media would offer the possibility to investigate regimes which are not fully modelled here. For example, obtaining an accurate description of the focusing of dispersive waves in the vicinity of the caustic remains an open problem. One could certainly imagine that the Gaussian beam approximation used here could be extended into the dispersive regime, but such an extension lies beyond the scope of this paper. Another interesting regime to explore, experimentally and theoretically, is the one arising from Berry's original proposal, namely the formation of non-linear waves as a result of the focusing process. As the wave focuses, its amplitude will locally increase. This amplification may result in a breakdown of the linear description studied in this paper and one may expect non-linear processes to come into play, such as the generation of solitary or rogue waves~\cite{Grimshaw2007,grimshaw_1970,PhysRevLett.106.204502}.

\subsection*{Acknowledgments}
SW acknowledges support provided by the Leverhulme Research Leadership Award (RL-2019-020), the Royal Society University Research Fellowship (UF120112) and the Royal Society Enhancements Grant (RGF/EA/180286 and RGF/EA/181015), and partial support by the Science and Technology Facilities Council (Theory Consolidated Grant ST/P000703/1), the Science and Technology Facilities Council on Quantum Simulators for Fundamental Physics (ST/T006900/1) as part of the Quantum Technologies for Fundamental Physics programme.
S.D.~acknowledges financial support from the Science and Technology Facilities Council (STFC) under Grant No.~ST/P000800/1, and from the European Union's Horizon 2020 research and innovation programme under the H2020-MSCA-RISE-2017 Grant No.~FunFiCO-777740.
T.T acknowledges financial support from STFC under the Quantum Technologies Grant No.~ST/T005858/1 .

\appendix
\section{Raychauduri's equation and Ricci tensor}\label{app:Ricci}
The full Raychauduri equation for our system is given by~\cite{Hawking:1973uf}
\begin{equation}\label{Raychaudhuri_equation}
    \frac{d\vartheta}{d\lambda} = -\vartheta^{2} - 2\sigma^{2} + 2\omega^{2} - R_{\mu\nu} k^{\mu} k^{\nu}
\end{equation}
Here $k^{\mu}$ is the tangent vector to the null geodesics associated to the expansion scalar $\vartheta$, $\omega$ is the vorticity scalar, $\sigma$ is the shear scalar and $R_{\mu \nu}$ the Ricci tensor calculated from the effective metric. The key difference between our case and the gravitational setting is the number of spatial dimensions. In the 2D setting, 
the Raychaudhuri equation is `missing' the factor of $1/2$ in front of $\vartheta^{2}$, and the  shear tensor vanishes identically. In addition, geodesics are hypersurface orthogonal ($k_\mu$ is a gradient) and so the vorticity tensor vanishes as well~\cite{Dempsey:2016wad}. Hence  Eq.~\eqref{Raychaudhuri_equation} reduces to
\begin{equation}\label{Analogue_Raychaudhuri}
    \frac{d\vartheta}{d\lambda} = -\vartheta^{2} - R_{\mu \nu} k^{\mu} k^{\nu} .
\end{equation}

From the (inverse) metric in Eq.~\eqref{inverse_metric}, we can explicitly compute the Ricci tensors, as follows. Since the metric is time-independent it implies that all time components of the Christoffel symbols vanish. Defining $f = \frac{1}{2h(r)r}\frac{dh}{dr}$, then the connection form ${A^{\mu}}_{\: \nu} = {\Gamma^{\mu}}_{\rho\nu}dx^{\rho}$ in $(t,x,y)$ coordinates (suppressing the time components) is:
\begin{equation}\label{connection_form}
    A = \begin{pmatrix}
    -xfdx - yfdy & yfdx + xfdy \\
    -xfdy + yfdx & -yfdy - xfdx
    \end{pmatrix}
\end{equation}
The curvature form is $F = dA + A \wedge A$, and $A \wedge A$ vanishes in our case, leaving $F = dA$ ($d$ is the usual exterior derivative, and the wedge product of two matrices is matrix multiplication with wedge products on each component). Consequently,
\begin{equation}
    F = dA = \begin{pmatrix}
    0 & (\partial_{x}(xf) + \partial_{y}(yf))dx\wedge dy \\
    -(\partial_{x}(xf) + \partial_{y}(yf))dx\wedge dy & 0
    \end{pmatrix} 
\end{equation}
A calculation, best illustrated with index notation, confirms that
\begin{equation}
    {F^{\mu}}_{\nu} = {F^{\mu}}_{\nu \rho\sigma} dx^{\rho} \wedge dx^{\sigma} = ({F^{\mu}}_{ \nu\rho\sigma} - {F^{\mu}}_{\nu\sigma\rho}) dx^{\rho} \otimes dx^{\sigma} = {R^{\mu}}_{ \nu\rho\sigma} dx^{\rho} \otimes dx^{\sigma}
\end{equation}
Performing a contraction of the $\mu$ and $\rho$ indices yields the Ricci tensor in the form
\begin{equation}\label{ricci_tensor}
R_{\mu\nu} dx^\mu dx^{\nu} = (\partial_{x}(xf) + \partial_{y}(yf))dx^2 + (\partial_{x}(xf) + \partial_{y}(yf)) dy^2.
\end{equation}

\section{Born approximation to the scattering cross section}\label{App:Born}
We follow here the approach of~\cite{PhysRevD.79.064014}.
To apply the Born approximation, we first rewrite the wave equation~\eqref{waveeq_mode} as:
\begin{equation}
\phi_{\omega,m}'' + c^{-2}(r)\left( \omega^2 - gh(r)\frac{(m^2 - 1/4)}{r^2}\right) \phi_{\omega,m} = 0.
\end{equation}
We then regroup the effect of the varying height in a potential $U$ such that the wave equation becomes:
\begin{equation}
    \frac{d^2\phi_{\omega,m}}{dx^2} + \left( \lambda^2 - \frac{(m^2 - 1/4)}{x^2} + U(x) \right) \phi_{\omega,m} = 0,
\end{equation}
where $x=r/r0$, $\lambda = \omega r_0/c_\infty$,  and the potential $U$ is given by:
\begin{equation}
    U(x) = \lambda^2 \left(\frac{c^2_\infty}{c^2(x)} - 1\right).
\end{equation}

This equation can be put into its integral form as:
\begin{equation}
\phi_{\omega,m}(x) = \bar{\phi}_{\omega,m} - \int_0^\infty{G(x,x_0)U(x_0) \phi_{\omega,m}(x_0) dx_0},
\end{equation}
where $\bar{\phi}_{\omega,m}$ is a solution to equation of~\eqref{born_we} with $U =0$, and $G(x,x_0)$ is the Green's function satisfying:
\begin{equation}\label{born_we}
\frac{d^2G}{dx^2} + \left( \lambda^2 - \frac{(m^2 - 1/4)}{x^2} \right) G(x) = -\delta(r-r_0).
\end{equation}
More explicitly, we have that:
\begin{equation}
G(x,x_0) = \left\{
            \begin{array}{ll}
                  x x_0\  j_{m-1/2}(\lambda x)y_{m-1/2}(\lambda x_0) \quad \text{for}\ x>x_0 
                  \\
                   x x_0\  j_{m-1/2}(\lambda x_0)y_{m-1/2}(\lambda x) \quad \text{for}\ x<x_0 
                \end{array}
              \right.
\end{equation}
where $j_m$ and $y_m$ are spherical Bessel functions of the first and second kind. Substituting the Green function into the integral form of the wave equation we get:
\begin{eqnarray}
\phi_{\omega,m}(x) &=& \bar{\phi}_{\omega,m} - \int_0^x{x x_0j_{m-1/2}(\lambda x_0) y_{m-1/2}(\lambda x)U(x_0) \phi_{\omega,m}(x_0) dx_0} \\
& &- \int_{x}^\infty{x x_0j_{m-1/2}(\lambda x) y_{m-1/2}(\lambda x_0)U(x_0) \phi_{\omega,m}(x_0) dx_0}, \nonumber
\end{eqnarray}

As $x\rightarrow \infty$, the third in the right-hand side is negligible and we therefore have that:
\begin{equation}
\phi_{\omega,m}(x) = \bar{\phi}_{\omega,m} - x y_{m-1/2}(\lambda x)\int_0^\infty{x_0j_{m-1/2}(\lambda x_0) U(x_0) \phi_{\omega,m}(x_0) dx_0}.
\end{equation}
The homogeneous solution, $\bar{\phi}_{\omega,m}$, is a plane wave. Using the asymptotic expansion of the Bessel's functions as well as the azimuthal decomposition of the plane wave, we get:
\begin{equation}\label{integral_form}
\phi_{\omega,m} = i^m\sqrt{\frac{2}{\pi\lambda}} \cos\left(\lambda x - \frac{m\pi}{2} - \frac{\pi}{4}\right) - \sqrt{\frac{2}{\pi\lambda}} \sin\left(\lambda x - \frac{m\pi}{2} - \frac{\pi}{4}\right)\int_0^\infty{x_0j_{m-1/2}(\lambda x_0) U(x_0) \phi_{\omega,m}(x_0) dx_0}.
\end{equation}

So far, the calculation is exact and no approximation has been applied. We now approximate the solution, $\phi_{\omega,m}$, by substituting $\phi_{\omega,m} \rightarrow \bar{\phi}_{\omega,m}$ in the integral form~\eqref{integral_form}. This gives for $r \rightarrow \infty$:
\begin{equation}
\phi_{\omega,m} \approx i^m\sqrt{\frac{2}{\pi\lambda}} \left[\cos\left(\lambda x - \frac{m\pi}{2}- \frac{\pi}{4}\right) - \sin\left(\lambda x - \frac{m\pi}{2}- \frac{\pi}{4}\right)I\right].
\end{equation}
with
\begin{equation}\label{integral_born}
I = \lambda\int_0^\infty{x_0^2 j_{m-1/2}(\lambda x_0)^2 U(x_0)  dx_0}.
\end{equation}
By comparing this form of the solution with the requirement that the wave be a superposition of a plane wave and a purely outgoing waves, we can express the phase shift in the Born approximation as:
\begin{equation}\label{born_phase_shift}
e^{2i\delta_m^{B}} = \frac{1+iI}{1-iI}.
\end{equation}
Since $I$ is real, we can see that $|e^{2i\delta_m^{B}}| = 1$.

Using the underwater island profile~\eqref{obstacle}, we can expand the propagation speed~\eqref{speed} as a power series in $1/x$:
\begin{equation}
\frac{1}{c^2(r)} \approx \frac{1}{c_\infty^2}\left[ 1 + \left( 1 - A\right)\frac{1}{x^n} + (A^2-A)\frac{1}{x^{2n}} + (A^2-A^3)\frac{1}{x^{3n}} + \mathcal{O}\left( \frac{1}{x^{4n}}\right)\right],
\end{equation}
where $A = c_0^2/c_\infty^2$ and expand the potential $U$ as: 
\begin{equation}\label{U_exp}
U(x) = \lambda^2\left[ (1-A)\frac{1}{x^n} + (A^2-A)\frac{1}{x^{2n}} + (a^2 - A^3)\frac{1}{x^{3n}} \right].
\end{equation}
Using this expansion with $n=4$ we can get the various contribution to the integral $I$, by using the properties of the spherical Bessel function~\cite{Stegun}:
\begin{equation}
I = I_1 + I_2 + ... 
\end{equation}
with
\begin{eqnarray}\label{born_large_m}
I_1 &=&  \lambda^4(1-A) \int_0^\infty{}u^{-2}j_{m-1/2}(u)^2du = \lambda^4(1-A) \frac{\pi}{8(m^3 - m)}.
\end{eqnarray}

\subsection{Born approximation to the deflection angle}
Using the Born approximation, we can estimate the deflection angle in the large m limit $m$.
Using~\eqref{born_phase_shift}, we deduce that the Born phase shift is given by:
\begin{equation}
\delta_m^B = \frac{1}{2}\arctan\left( \frac{2I}{1-I^2}\right).
\end{equation}
In the large $m$ limit, we have that $I \approx I_1$ and $I \ll 1$. Using the Taylor expansion of $\arctan$ near 0, we have that:
\begin{equation}
\delta_m^B \approx I_1.
\end{equation}

Finally using the relation between the deflection angle and the phase shift $\Theta = 2 d \delta_m/dm$, we can estimate the deflection angle in the Born approximation as:
\begin{equation}
\Theta = 2 \frac{d I_1}{dm}.
\end{equation}

\subsection{Application to the scattering cross-section}

The Born approximation gives us a way to estimate the phase-shift by evaluating the integral~\eqref{integral_born}. The full potential $U$ can be used in the integral or one can substitute the leading order expansion of $U$ for large $x$ in order to analytically get the large $m$ limit to the phase-shift. These approximations can be used in combination with the numerical method detailed in \ref{sec:numerics} to accurately compute the scattering cross-section.

The scattering cross-section can be computed from the partial wave expansion by evaluating the infinite sum given in Eq.\eqref{scattering_cs}. This sum can be split into three component which will be evaluated using different method:
\begin{equation}
    f_\omega(\theta) = \sum_{m=0}^{N_1} a_m \cos(m\theta) + \sum_{m=N_1+1}^{N_2} a_m \cos(m\theta) + \sum_{m=N_2+1}^{\infty} a_m \cos(m\theta) = S_1 + S_2 + S_3.
\end{equation}

$S_1$, corresponding the low-$m$ contribution, can be evaluated by using the phase-shift coefficients obtained by solving numerically the wave equation.

$S_2$, corresponding the mid-$m$ contribution, can be evaluated the phase shifts via the Born approximation by numerically integrating~\eqref{integral_born} and using the full potential $U$. 

$S_3$, corresponding the large-$m$ contribution, can be estimated by using the large-$m$ limit analytic expression for the phase-shifts in the Born approximation with Eq.\eqref{born_large_m}. Indeed $S_3$ is given as a combination of the Hurwitz-Lerch transcendent $\Phi(z,s,a)$ which satisfy:
\begin{equation}
   \Phi(z,s,a) = \sum_{k=0}^{\infty} \frac{z^k}{(k+a)^s}.
\end{equation}
By including high order terms in the large-$m$ limit of the Born phase shifts, one can increase the accuracy of the estimate for $S_3$.

\begin{figure}[!h]
 \includegraphics{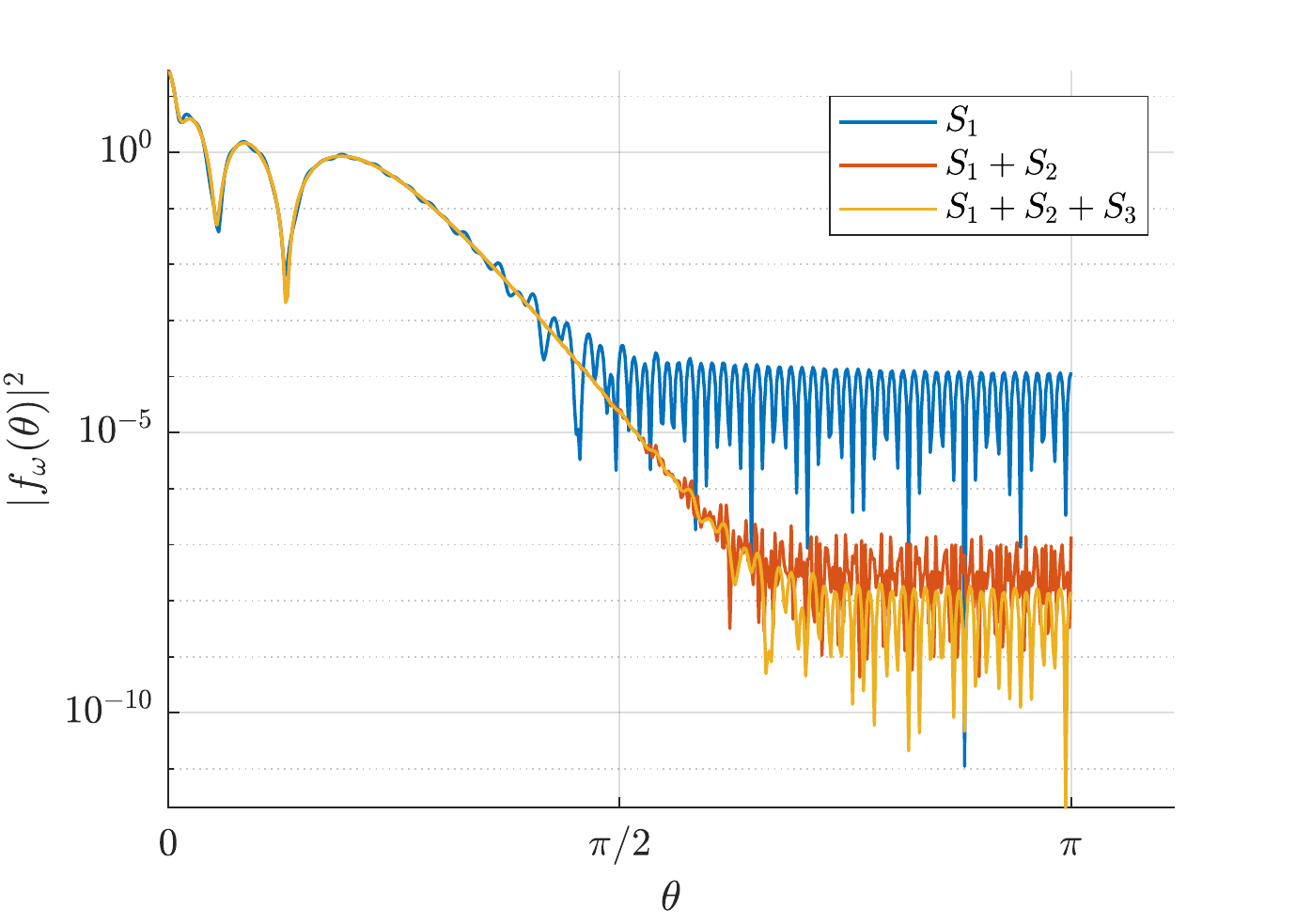}
 \caption{Contribution to the scattering cross section for $f=\frac{\omega}{2\pi} = 8$Hz of the various sums $S_1$, $S_2$ and $S_3$. Here the cut off between the sums are $N_1 = 80$ and $N_2=280$.}
 \label{fig:scat_cross_app}
\end{figure}

Figure \ref{fig:scat_cross_app} shows the contribution of the different sums in computing the scattering cross section. We can see that the biggest gain in accuracy comes from including $S_2$ which consists in approximating the phase shifts using the Born approximation with the full potential $U$.

\section{Parabolic submerged island}\label{App:parabola}
Here we consider the case of a finite size obstacle which has a parabolic form.
The height profile is represented by
\begin{equation}
h(r) = \left\{
                \begin{array}{ll}
                  h_\infty \quad \text{for}\ r\geq R\\
                  h_0 - Br^2 \quad \text{for}\ r\leq R
                \end{array}
              \right.
\end{equation}
with $B =(h_0 - h_\infty)/R^2$.
\subsection{Region I : $r \leq R$}
In the region $r \leq R$, the wave is propagating over the parabola. The wave equation is this region reduces to
\begin{equation}
g(h_0 - Br^2)\phi_{\omega,m}'' + \left( \omega^2 - g(h_0 - Br^2)\frac{(m^2 - 1/4)}{r^2}\right) \phi_{\omega,m} = 0.
\end{equation}
which can be rewritten as:
\begin{equation}
(1 - br^2)\phi_{\omega,m}'' + \left( k_0^2 - (1 - br^2)\frac{(m^2 - 1/4)}{r^2}\right) \phi_{\omega,m} = 0,
\end{equation}
where $k_0 = \omega_0/c_0$, $c_0^2 = gh_0$ and $b = B/h_0$. This equation is a hypergeometric equation and its solutions are given by hypergeometric function ${}_{2}F_{1}$:
\begin{equation}
\phi_{\omega,m} = A_1 r^{m+1/2}{}_{2}F_{1}\left( \frac{bm - s}{2b},\frac{bm + s}{2b},1+m,br^2\right) + A_2 r^{1/2-m}{}_{2}F_{1}\left( \frac{-bm + s}{2b},-\frac{bm + s}{2b},1-m,br^2\right),
\end{equation}
with $s = \sqrt{b(k_0^2 + bm^2)}$.
From the asymptotic expression of the hypergeometric functions when $r \rightarrow 0$, and the boundary condition at the origin, we deduce that $A_2 = 0$. 
Therefore the physical solution is:
\begin{equation}
\phi_{\omega,m} = A_1 r^{|m|+1/2}{}_{2}F_{1}\left( \frac{b|m| - s}{2b},\frac{b|m| + s}{2b},1+|m|,br^2\right).
\end{equation}
\subsection{Region II : $r\geq R$}

In the region $r \geq R$, the speed of the waves is a constant and the wave equation has the form
\begin{equation}
c_\infty^2\phi_{\omega,m}'' + \left( \omega^2 - c_\infty^2\frac{(m^2 - 1/4)}{r^2}\right) \phi_{\omega,m} = 0.
\end{equation}
The solutions to his equation are given in terms of Bessel functions:
\begin{equation}
\phi_{\omega,m} = \sqrt{r}\left(B_1 J_m(k_\infty r) + B_2 Y_m(k_\infty r)\right). 
\end{equation}
From the asymptotic expansion of the Bessel functions, we can expression the in and out coefficient in terms of $B_1$ and $B_2$ as
\begin{eqnarray}
A^{\text{out}}_{\omega,m} &=& \frac{B_1 -i B_2}{\sqrt{2\pi k_\infty}} e^{-i\pi/4} (-i)^m \\
A^{\text{in}}_{\omega,m} &=& \frac{B_1 +i B_2}{\sqrt{2\pi k_\infty}} e^{i\pi/4} (i)^m 
\end{eqnarray}
This implies that the phase shift can be simply written as
\begin{equation}
e^{2i\delta_m} = \frac{B_1 -i B_2}{B_1 + i B_2}.
\end{equation}

\subsection{Matching condition}
The solutions in region I and region II should agree at $r = R$. This implies that:
\begin{equation}\label{match1}
A_1 R^{|m|}\mathcal{F} = B_1\mathcal{J}_m + B_2 \mathcal{Y}_m
\end{equation}
where $\mathcal{F} = {}_{2}F_{1}\left( \frac{b|m| - s}{2b},\frac{b|m| + s}{2b},1+|m|,bR^2\right)$,
 $\mathcal{J}_m = J_m(k_\infty R)$ and $\mathcal{Y}_m = Y_m(k_\infty R)$.

Similarly, their derivatives must also be continuous. This implies that
\begin{eqnarray}\label{match2}
\frac{A_1}{2}R^{|m|-1/2}\left\lbrace(1+2|m|)\mathcal{F} - R^2\frac{k_0^2}{(1+|m|)}\mathcal{F}_2 \right\rbrace = &\frac{1}{2\sqrt{R}}&\left\lbrace B_1\left(k_\infty R (\mathcal{J}_{m-1}-\mathcal{J}_{m+1}) + \mathcal{J}_m\right) \right. \nonumber \\
&+& \left. B_2\left(k_\infty R (\mathcal{Y}_{m-1}-\mathcal{Y}_{m+1}) + \mathcal{Y}_m\right) \right\rbrace,
\end{eqnarray}
with
\begin{equation}
\mathcal{F}_2 = {}_{2}F_{1}\left[ \frac{1}{2}\left(2+|m| - \frac{s}{b} \right), 1 + \frac{b|m|+s}{2b},2+m,b R^2\right].
\end{equation}
The two matching conditions gives us a relation between $B_1$ and $B_2$:
\begin{equation}
B_1 = \alpha B_2.
\end{equation}
where the coefficient $\alpha$ is fully determined by the parameters of the system.
Explicitly
\begin{equation}
\alpha = \frac{\nu a_1 - \mathcal{Y}_m a_2}{\mathcal{J}_ma_2 - \gamma a_1},
\end{equation}
with
\begin{eqnarray}
a_1 &=& R^{|m|}\mathcal{F} \\
a_2 &=& \frac{R^{|m|-1/2}}{2}\left( (1+2|m|)\mathcal{F} - R^2 \frac{k_0^2}{1+|m|}\mathcal{F}_2  \right) \\
\gamma &=& \frac{1}{2\sqrt{R}}\left( k_\infty R(\mathcal{J}_{m-1} - \mathcal{J}_{m+1}) + \mathcal{J}_m\right) \\
\nu &=& \frac{1}{2\sqrt{R}}\left( k_\infty R(\mathcal{Y}_{m-1} - \mathcal{Y}_{m+1}) + \mathcal{Y}_m\right)
\end{eqnarray}
Inserting this into the expression for the phase shift, we get that
\begin{equation}
e^{2i\delta_m} = \frac{\alpha -i}{\alpha +i}.
\end{equation}

\bibliographystyle{apsrev4-1}
\bibliography{bibli}

\end{document}